\documentclass[11pt,x11names,a4paper]{article}
\pdfoutput=1

\usepackage[utf8]{inputenc}
\usepackage{lmodern}
\usepackage[T1]{fontenc} 
\usepackage{microtype} 

\usepackage[a4paper, left=25mm, right=25mm, top=30mm, bottom=25mm]{geometry} 
\usepackage{fancyhdr} 
\usepackage{abstract}

\usepackage{titlesec} 
\titleformat{\section}[block]{\large\bfseries\centering}{\thesection}{1em}{} 
\titleformat{\subsection}[block]{\bfseries}{\thesubsection}{1em}{} 

\usepackage{xcolor}

\usepackage{cite}
\usepackage{hyperref}
\hypersetup{
	colorlinks=true,
	linkcolor=Blue4,
	citecolor=Red4,
	urlcolor=Green4,
	linktoc=page
}

\usepackage{enumerate}
\usepackage{graphicx}
\usepackage[percent]{overpic}
\usepackage{tikz}
\usepackage{subcaption}
\usepackage[many]{tcolorbox}
\usepackage{booktabs}
\tcbset{highlight math style={enhanced,
  colframe=DeepSkyBlue4,colback=Honeydew1}}

\usepackage{amsmath,amssymb,slashed,mathbbol,mathtools}
\usepackage{amsthm}
\numberwithin{equation}{section}
\usepackage{slashed}

\DeclareSymbolFontAlphabet{\amsmathbb}{AMSb}%

\newcommand{\dd}{\mathrm{d}}

\newcommand{\e}{\mathrm{e}}
\newcommand{\rmi}{{i}}
\newcommand{\rme}{\mathrm{e}}
\newcommand{\rmd}{\mathrm{d}}
\newcommand{\w}{\wedge}

\newcommand{\be}{\begin{equation}}
\newcommand{\ee}{\end{equation}}
\newcommand{\bea}{\begin{eqnarray*}}
\newcommand{\eea}{\end{eqnarray*}}

\renewcommand{\i}{\iota}
\newcommand{\f}[2]{\frac{#1}{#2}}
\newcommand{\R}{\mathbf{R}}
\newcommand{\C}{\mathbf{C}}
\newcommand{\Z}{\mathbf{Z}}

\newcommand{\vol}{\text{vol}}

\newcommand{\Tr}{\text{Tr}~}
\newcommand{\trace}{\Tr}

\newcommand{\CP}{\C{P}}
\newcommand{\RP}{\R{P}}

\newcommand{\SU}{\mathop{\rm SU}}
\newcommand{\SO}{\mathop{\rm SO}}

\newcommand{\U}{\mathop{\rm {}U}}

\newcommand{\OSp}{\mathop{\rm {}OSp}}

\newcommand{\Yslm}[1]{\,{}_{#1}Y_{lm}}

\title{\fontsize{22pt}{26pt}\selectfont\textbf{Quantized Strings and Instantons\\ in Holography}\vspace{2mm}}

\author{
\large{
\href{mailto:ffg@hi.is}{Fri{\dh}rik Freyr Gautason},$^1$ \href{mailto:vgmp@hi.is}{Valentina Giangreco M. Puletti},$^1$ and \href{mailto:jvanmuid@sissa.it}{Jesse van Muiden}$^{2,3}$}\\[5mm]
{}$^1${\normalsize Science Institute, University of Iceland}\\
{\normalsize Dunhaga 3, 107 Reykjav{\'i}k, Iceland}\\[5mm]
{}$^2${\normalsize SISSA}\\
{\normalsize Via Bonomea 265, I 34136 Trieste, Italy}\\[5mm]
{}$^3${\normalsize INFN - Sezione di Trieste}\\
{\normalsize Via Valerio 2, I-34127 Trieste, Italy}
}

\date{}

\begin{document}  
{\hypersetup{urlcolor=black}\maketitle}
\thispagestyle{empty}

\begin{abstract}
\noindent 
We study worldsheet instantons in holographic type IIA backgrounds directly in string theory. The first background is a dimensional reduction of AdS$_7\times S^4$ and is dual to the maximally supersymmetric Yang-Mills theory on $S^5$. The second background is AdS$_4\times \CP^3$ dual to ABJM in the type IIA limit. We compute the one-loop partition function of the fundamental string in these backgrounds and show that the result is in exact agreement with field theory predictions. 
We argue that for higher rank instantons, the string partition function takes a product form of the single instanton partition function times the contribution of two orbifolds on the worldsheet. We determine the orbifold factor to be $n^{-3/2}$ where $n$ is the instanton rank. 
With this result, we reproduce the series of non-perturbative corrections in $\alpha'$ to the planar $S^5$ free energy. 
When studying the worldsheet instanton partition function on $\CP^3$, we encounter twelve fermionic and twelve bosonic zero modes. 
By deforming the ABJM theory, the zero-modes are lifted and consequently the tower of worldsheet instantons can be evaluated and matched to known results in the QFT. 
 As a by-product, we determine a series of higher rank instanton corrections to the free energy of the mass-deformed and orbifolded ABJ(M) theory.
\end{abstract}

\newpage

\setcounter{tocdepth}{2}
\tableofcontents


\section{Introduction}
\label{sec:introduction}
As a strong-weak duality, holography has proven an invaluable tool in the exploration of both strongly coupled quantum field theories and quantum gravity. 
We will use the remarkable progress made in computing exact answers for observables in strongly coupled quantum field theories, to learn about gravity beyond the classical limit. One of the main tools that allowed for such progress on the QFT side is supersymmetric localization, which was originally used in its modern form by Pestun to prove a conjectured form of the expectation value of $\frac12$-BPS Wilson loop in ${\cal N}=4$ supersymmetric Yang-Mills theory in four dimensions \cite{Erickson:2000af,Drukker:2000rr,Pestun:2007rz}. Since then, supersymmetric localization has been utilized in a variety of situations for theories with different amount of supersymmetry and other global symmetries to a great success (see \cite{Pestun:2016zxk} for a review).

These direct QFT results have permitted numerous precision tests of holography, which maps the strong coupling regime of the QFT to the semi-classical regime of string or M-theory. Going beyond leading order in the expansion parameter, $L/\ell_{s}\gg 1$ where $L$ is a characteristic length scale of the ten-dimensional geometry and $\ell_{s}$ is the higher dimensional Planck length,\footnote{The higher dimensional Planck length is identified with the string length (and thus $\alpha'$) in ten dimensions.} has turned out to be a challenge since higher orders in the expansion require the precise knowledge of higher derivative corrections to the gravitational theory which are generally out of reach. 

In principle, the method for determining the higher derivative corrections to ten- or eleven-dimensional gravity is straight-forward. First, (super)symmetry is used to restrict the form of the higher derivative terms at a given order to a manageable number. The coupling constants of these supersymmetric invariants are at this stage still undetermined and have to be fixed through physical observables, e.g. graviton scattering or beta functions on the worldsheet. In practice, however, especially the first step is technically rather unfeasible already at the first $\alpha'$ correction, let alone at higher orders.

We propose an alternative strategy to make a direct contact between string theory and quantum field theory. Rather than using string theory to compute higher derivative corrections to the supergravity theory, we will consider quantizing string theory in a given curved background. Of course the well-known difficulties of quantizing string theory in RR backgrounds are still persistent, nevertheless we argue this to be a worthwhile endeavor due to the by now many impressive quantum field theory results, which can be used as a secure holographic anchor to hold on to. 

In this paper we will focus on one instance where string theory can be analyzed perturbatively in a curved background with RR fluxes turned on. Explicitly, we will consider worldsheet instanton contributions to the type IIA string partition function in a variety of backgrounds. These contributions are non-perturbative in $L/\ell_{s}$ but at leading order in the string genus expansion. To this end we will use the Green-Schwarz (GS) formulation of string theory which has proven useful in similar contexts in past holographic studies. 

Worldsheet instantons in string theory, in the context of holography, have been analyzed in the past at the leading order, i.e. the classical string action has been computed for a stable configuration of the string. However, to the best of our knowledge, a computation of the one-loop fluctuations around a string configuration in this context has not been carried out before. The main reason is that dealing with various Jacobi and measure factors in the Green-Schwarz formulation has proven challenging. As emphasized in \cite{Drukker:2000ep,Giombi:2020mhz}, these measure factors are universal but depend on regularization scheme in such a way that the one-loop partition function of the string is well defined and scheme independent. This is where holography can help us, namely, we fix these measure factor ambiguities by comparing with exact results from field theory. Once this has been done in one background, the same value of this measure factor can be used in other contexts and should not have to be adjusted as long as the regularization scheme and topology of the worldsheet remain the same. We verify this expectation by comparing a string theory computation to a field theory prediction in a few non-trivial examples. 

The first example studied in this paper is five-dimensional supersymmetric Yang-Mills theory on $S^5$. This theory was localized in \cite{Kim:2012ava} and the partition function could be computed explicitly including Yang-Mills instanton effects \cite{Kim:2012qf}. We will focus on the large $N$ limit of this theory where the field theory answer for the $S^5$ partition function takes a particularly simple form. Specifically, the answer has three terms that are perturbative in the coupling constant as well as infinitely many non-perturbative terms. The holographic dual geometry is a dimensional reduction of AdS$_7\times S^4$ to type IIA supergravity \cite{Bobev:2018ugk}, where the coupling constant of the field theory is mapped to a length scale of the ten-dimensional geometry. As previously mentioned, our focus is the series of non-perturbative terms in $L/\ell_{s}$, given by worldsheet instantons, which on the field theory side is mapped to a series of non-perturbative terms in the coupling constant. By carefully studying the first term in the series, we fix the measure factor ambiguity in the string partition function in such a way that an exact match is obtained. We then also consider higher instanton terms corresponding to the remaining non-perturbative corrections to the $S^5$ partition function. On the string theory side, this culminates in evaluating the partition function of a multi-wound string, i.e. the string is specified by a degree $n$ map from the worldsheet $S^2$ to the target-space $S^2$ it wraps. A very similar setup has been studied extensively in the context of higher rank Wilson loop holography, but all attempts at a match have so far failed. We argue that the multi-wound partition function of a string is related to the singly wound partition function times a contribution associated to the orbifold singularities on the worldsheet. This allows us to conjecture that the multi-wound partition function is simply a rescaling of the single string partition function by a multiplicative factor due to each orbifold point. We fix this multiplicative factor by revisiting the multi-wound string in AdS$_5\times S^5$ which is dual to higher rank Wilson loop in ${\cal N}=4$ SYM. With this conjecture at hand and the explicit multiplicative factor, we can reproduce the entire series of non-perturbative terms in the supersymmetric partition function of maximal SYM on $S^5$.

The second example we study is the ABJM theory and its various extensions. Now, all the measure factors have been fixed in our previous computation and so this example serves as an important validation of our procedure. After a brief review of the computation of instanton corrections to the $S^3$ partition function on the field theory side, we indeed verify that our string theory computation reproduces the expected answer. The structure of the computation is, however, completely different compared to our first example, and turns out to be much more intricate. First, in contradiction to na\"ive expectation, we find that both strings and anti-strings contribute to the supersymmetric partition function. 
This is allowed since the AdS$_4\times \CP^3$ geometry dual to ABJM theory does not have the three-form field strength turned on, which would affect the stability of either the string or the anti-string. Second, we find twelve fermionic and twelve bosonic zero-modes in the one-loop spectrum of the string. In fact, the zero-modes had been observed in previous work \cite{Cagnazzo:2009zh}. Naively, one would expect that the fermionic zero-modes would render the string partition function trivial, and only observables for which the zero-modes have been soaked up by operators would be non-trivial. However, since we know that the field theory answer for the instanton partition function is finite \cite{Drukker:2010nc}, we argue that this is not the case and instead that the bosonic and fermionic zero-modes conspire to result in a finite valued string partition function. To evaluate the finite answer, we lift the zero-modes through a supersymmetric deformation of the background. The most important feature of our computation is that the string finds not one stable configuration once the zero-modes are lifted, but two. Hence, the string (and the anti-string) contributes twice the expected answer. This is reminiscent of supersymmetric localization of quantum field theories, and indeed we expect a careful argument could be made to demonstrate the localization of the string. As mentioned, combining all terms we again find  exact agreement with the field theory. For the higher instantons we again apply the conjectured form of the multi-wound worldsheet instanton partition functions and find that it correctly reproduces a tower of higher instantons known in the QFT. Not all known QFT instantons are reproduced, however, due to the fact that we neglect D2-brane instantons and worldsheet instantons wrapping more complicated cycles in the $\mathbf{C}P^3$ background.

The deformation used to lift the zero-modes is interesting in its own right. It is a supersymmetric mass deformation of the ABJM theory studied in \cite{Jafferis:2011zi,Freedman:2013ryh}. Supersymmetric localization has been utilized to give the perturbative answer for the free energy, but as far as we are aware, instanton corrections in the type IIA limit have not been studied before.\footnote{From the QFT point of view instantons were studied in the small $k$ expansion for a specific mass-deformed ABJM theory in \cite{Nosaka:2015iiw}.} Using our results, we give a prediction for a tower of instanton corrections to the mass deformed ABJM theory and its ABJ cousin.

The final example we study is an orbifold of the ABJM theory in the IIA limit \cite{Honda:2014ica}. The orbifold acts on the $\CP^3$ in a non-trivial way such that orbifold singular points are visible to the worldsheet instantons. Given our conjecture for orbifold points, we are able to derive the first instanton contribution to the free energy of this theory and we find an exact match with field theory predictions \cite{Hatsuda:2015lpa}.

The paper is organized as follows. In section \ref{sec:GS}, we review and discuss the Green-Schwarz formulation of string theory and describe how the string partition function is related to the free energy of the dual QFT. We discuss the spectrum of string fluctuations for a spherical worldsheet and how to compute the corresponding one-loop partition function. We also discuss in general terms, how higher instantons contribute and our expectation for their partition function based on the analogy with multi-wound strings in AdS$_5\times S^5$. In section \ref{sec:5dsym}, we discuss five-dimensional super Yang-Mills on $S^5$. We review results from localization before studying string theory in the dual geometry. After computing the string partition function for the single string, we use the localization result to fix our measure factor ambiguity (denoted by $C(2)$ in this paper). We then discuss higher worldsheet instantons, i.e. multi-wound strings, in this geometry and predict their partition functions using our conjecture from section \ref{sec:GS}. This allows us to compute from the string theory perspective the entire series of instanton corrections to the field theory free energy and match with the localization prediction. In section \ref{sec:ABJM} we discuss a similar computation for the ABJ(M) theory. On the string theory side we note the presence of zero-modes and anticipate the finite zero-mode partition function (which we compute explicitly in section \ref{sec:3dCS}) that we can use to fully determine a tower of worldsheet instanton partition functions and match with field theory. In section \ref{sec:3dCS}, we discuss mass deformations of the ABJM theory that can be used to evaluate the zero-mode partition function in the undeformed theory, and additionally we study the orbifolded ABJM theory. We can then also give a prediction for the instanton corrections to the free energy of these two ABJM variants. Finally, in section \ref{sec:discussion} we summarize our results and discuss future work. We also include three appendices. In appendix \ref{MeasureFactors} we include a general discussion of measure factors in string theory. In appendix \ref{App: Fermionic monopole harmonics} we review the spectrum and eigenfunctions of the fermionic operators encountered in this paper. In appendix \ref{App: Holographic duals to mass deformed ABJM} we give the eleven-dimensional geometry dual to ${\cal N}=4$ mass deformation of ABJM, and also another type IIA solution that is dual to the $\SU(3)$-deformation of ABJM theory which is an uplift of a solution from \cite{Freedman:2013ryh}.

 \section{Worldsheet instantons in the Green-Schwarz formulation}
 \label{sec:GS}

As discussed in the introduction, in this paper we compute worldsheet instanton corrections to the string partition function. The string partition function can be evaluated in the long wavelength approximation where it is dominated by saddle points
\begin{equation}\label{saddlepointexpansion}
{\cal Z}_\text{string} \approx \sum_\text{saddles} \e^{-S_\text{cl}}Z_\text{1-loop}\,,
\end{equation}
where $S_\text{cl}$ is the string action evaluated on the saddle point surface $M$ and simply equals its dimensionless area ${\cal A}$ plus the integral of the Kalb-Ramond two-form $B_2$,
\be\label{classicalAction}
S_\text{cl} = {\cal A} + \f{i}{2\pi\ell_s^2}\int B_2\,,\qquad {\cal A} = \f{1}{2\pi\ell_s^2} \int_M \vol_\gamma\,.
\ee
Here $\gamma_{ij}$ denotes the pull-back of the ten-dimensional metric to the worldsheet $M$. The factor of $i$ in the two-form term is due to the fact that we consider Euclidean worldsheets.

In the saddle point expansion \eqref{saddlepointexpansion}, the leading saddle is given by a `point-like' string, i.e. where the string embedding is trivial in spacetime. The classical action in this case is zero and the entire contribution comes from the one-loop action. Due to the trivial embedding, the one-loop theory has ten bosonic zero-modes corresponding to the collective motion of the string in the ten-dimensional ambient space. These zero-modes must be integrated over with the correct measure factor which is determined by the ten-dimensional target space metric. The remaining contribution to the one-loop partition function was shown to give minus the ten-dimensional Lagrangian \cite{Fradkin:1984pq,Fradkin:1985ys,Fradkin:1985fq} of supergravity\footnote{Strictly speaking the supergravity Lagrangian is not directly related to the string partition function of the trivial saddle at genus zero, but instead is related to its derivative with respect to a UV cutoff \cite{Tseytlin:1988tv,Tseytlin:1988rr,Tseytlin:2006ak} (see also \cite{Ahmadain:2022tew}). We expect this subtlety will only affect the relative coefficient between the two terms in the expression \eqref{Zstring} which in this paper will be fixed using holography as explained below. We thank Arkady Tseytlin for a discussion on this point.} and so we can write \eqref{saddlepointexpansion} as
\begin{equation}\label{Zstring}
{\cal Z}_\text{string} \approx - S_\text{sugra}+ \sum_\text{instantons}\e^{-S_\text{cl}}Z_\text{1-loop}\,,
\end{equation}
where the instanton sum runs over non-trivial saddles, i.e. ones with finite classical action. In holography, the suitably regularized on-shell supergravity action should reproduce the free energy of the dual QFT. Including worldsheet instantons, the string partition function should therefore reproduce minus the free energy of the dual QFT. In the saddle point expansion, this implies the holographic relation
\begin{equation}\label{Eq:Holography}
\tcbhighmath{~F_\text{QFT} \approx S_\text{sugra}- \sum_\text{instantons}\e^{-S_\text{cl}}Z_\text{1-loop}\,.~}
\end{equation}
In this equation we are working to leading order in the string coupling constant $g_s$, and to leading order in $L/\ell_s$ around each saddle. For the rest of the paper we will not consider the leading order contribution from the supergravity action, but instead discuss the sum over worldsheet instantons. To avoid confusion we emphasize that throughout the paper we will compare QFT free energies $(F)$ to string partition functions $(Z)$.

The one-loop partition function, $Z_\text{1-loop}$, receives contributions from three parts. The first contribution comes from the classical evaluation of the Fradkin-Tseytlin (FT) action~\cite{Fradkin:1984pq,Fradkin:1985fq}
\begin{equation}
S_\text{FT} = \f{1}{4\pi}\int_M   \Phi R_\gamma\vol_\gamma\,.
\end{equation}
The FT term contributes at one-loop due to the fact that it scales like $(L/\ell_s)^0$ where $L$ is the length scale of the geometry in question. This should be contrasted to the classical action which scales as $(L/\ell_s)^2$.
The second contribution is the actual one-loop partition function of the two-dimensional CFT living on the worldsheet. For the GS string this consists of ten non-interacting scalars and eight non-interacting fermions. Next, we have the contribution of string ghosts, which cancel the longitudinal fluctuations of the string reducing the physical scalar modes from ten to eight. The cancellation leaves a remainder which is expected to be universal (in the sense that we are going to explain below) and to only depend on the Euler characteristic $\chi$ of the worldsheet (see \cite{Drukker:2000ep,Giombi:2020mhz} for more discussion). We denote this term by $C(\chi)$.
Collecting all the terms we can write
\begin{equation}\label{eq:1-loop pf def}
Z_\text{1-loop}=\e^{-S_\text{FT}} C(\chi)(\text{Sdet}'{\mathbb{K}})^{-1/2}Z_\text{zero-modes}\,,
\end{equation}
where $\mathbb{K}$ collectively denotes the one-loop operators for the eight physical scalar and eight fermionic modes. When evaluating the partition function, zero-modes need to be treated separately, and their contribution is collected in $Z_\text{zero-modes}$, whereas the prime on $\text{Sdet}$ indicates that we exclude zero-modes when evaluating the functional determinants (see appendix \ref{MeasureFactors}).

The one-loop determinant $(\text{Sdet}'{\mathbb{K}})^{-1/2}$ diverges but the coefficient of the divergence is expected to be universal and to be controlled by the Euler characteristic \cite{Drukker:2000ep,Giombi:2020mhz}. The specific value of the finite term in $(\text{Sdet}'{\mathbb{K}})^{-1/2}$ depends of course on the regularization scheme used. Therefore, the precise numerical factors in the universal term $C(\chi)$ similarly depend on the regularization scheme, but in such a way that the product in \eqref{eq:1-loop pf def} is scheme independent. In this paper, we exclusively use $\zeta$-function regularization as explained in more detail in the next subsection. It is convenient to compute the determinants of rescaled operators for which the dependence on the length scale of the geometry $L$ has been eliminated. These are the operators on $M$ for which the radius has been set to one. 
This length scale has to be reintroduced at the level of the universal factor $C(\chi)$.%
\footnote{For comparison, we note that in \cite{Cagnazzo:2017sny,Medina-Rincon:2018wjs,Gautason:2021vfc} a different regularization scheme was used, hence the value of $C(1)$ there cannot be directly compared to the value reported here in \eqref{C1eq}.}
For this reason we expect%
\footnote{For unrescaled operators, the partition function scales as $-\f12 \log \text{Sdet}'{\mathbb{K}} \sim \chi \log L\sim \f\chi{2}\log{\cal A}$.}
\be
C(\chi) \sim {\cal A}^{\chi/2}\,,
\ee
but fixing the numerical factor requires a careful analysis of all measure factors in the GS string path integral and assigning a finite value to some possible diverging factors.
Just as in \cite{Giombi:2020mhz}, we will not take this route, but instead use holography to fix the numerical factor. For a disk partition function, the constant was fixed in \cite{Giombi:2020mhz} by comparison to the 1/2-BPS Wilson loop in ${\cal N}=4$ SYM to be\footnote{In \cite{Giombi:2020mhz} the constant was given as $\sqrt{T/2\pi}$ where $T$ is the `effective tension' of the string in the geometry and is related to the dimensionless area by ${\cal A} = -2\pi T$.}
\be\label{C1eq}
C(1) = \f{\sqrt{-\cal A}}{2\pi}\,.
\ee
This was then also shown to be consistent with the evaluation of the vev of the 1/2-BPS Wilson loop operator in the ABJM theory.   In this paper we focus on sphere partition functions and therefore we need the value $C(2)$.  In section \ref{sec:5dsym} we use exact results for the 5D SYM partition function which include terms non-perturbative in the coupling to fix the coefficient $C(2)$ by evaluating the worldsheet instanton directly in the string dual. In this way we fix $C(2)$ to be
\be\label{C2eq}
C(2)= \f{{\cal A}}{8\pi^2}\,.
\ee
In section \ref{sec:ABJM} we compute worldsheet instanton corrections to the $S^3$ partition function of the ABJM theory and find a match with a matrix model computation, thus finding a non-trivial confirmation of the value \eqref{C2eq}.

\subsection{The one-loop action}
\label{ssec:oneloop}
When evaluating the one-loop partition function for the worldsheet instanton string we have to accurately determine the kinetic operators for the small fluctuations around the classical solution. This is done by expanding the Green-Schwarz action \cite{Cvetic:1999zs,Wulff:2013kga} to second order in the fields, i.e. the eight transverse scalar modes as well as the eight fermion directions. We refer the reader to \cite{Callan:1989nz,Drukker:2000ep,Forini:2015mca} for a detailed discussion. 

For the instantons encountered here, the worldsheet is spherical and equipped with the round metric,
\begin{equation}\label{sphericalworldsheet}
\dd s^2 = L^2 \dd \Omega^2_2 = L^2 \Big(\dd\theta^2 + \sin^2\theta \dd \varphi^2\Big)\,,
\end{equation}
where $L$ is some length scale set by the parent geometry. Due to the symmetry preserved by the instantons in these cases, the kinetic operators for the scalar and fermionic modes are restricted to be particularly simple. We can write them as follows\footnote{In our conventions $\slashed{\partial} = \sigma_1\partial_\theta + \sigma_2\partial_\varphi$ where $\sigma_{1,2,3}$ are the standard Pauli matrices.}
\begin{equation}\label{operatordef}
\begin{split}
{\cal K} = -D^2 + M_b^2\,,&\qquad \text{for scalars}\,,\\
{\cal D} = i\slashed{D} + m_1 \sigma_3 + m_2 \,,&\qquad\text{for fermions}\,. 
\end{split}
\end{equation}
Here $D_\mu = \nabla_\mu - i q A_\mu$ where $A_\mu$ is a monopole gauge field on $S^2$ such that the field strength $F=\dd A$ is normalized to have unit flux,
\begin{equation}
\f{1}{2\pi} \int_{S^2} F = 1\,,
\end{equation}
and $q$ is the integer charge of the field in question. Explicitly, the monopole gauge field takes the form
\begin{equation}\label{eq:monopolefield}
A = \f12(1-\cos\theta)\,\dd \varphi\,,\qquad F= \f12 \sin\theta \,\dd\theta\w \dd\varphi\,.
\end{equation}
The monopole field can arise as a result of a non-trivial $B_2$ field in the ten-dimensional geometry or if the normal bundle is non-trivially fibered over the worldsheet.%
\footnote{Semi-classical string quantization around a classical solution in the presence of a $B$-field is reviewed in \cite{Callan:1989nz}.}
We will encounter charged operators for the ABJM example studied in section \ref{sec:ABJM}. In that case we will have equally many fields with positive and negative charge such that charge conjugations is a symmetry of our theory. Reflection around the `time' circle $\varphi$ is similarly a symmetry and so, if we let ${\bf C}$ be the charge conjugation operator and ${\bf T}$ the $\varphi$-reflection operator, then we can define $\tilde {\cal D} = -\sigma_2 {\bf TC}{\cal D}\sigma_2$ and
\begin{equation}
\tilde {\cal D}{\cal D}=-D^2 + \f{R}{4} + M_f^2 + i q\slashed{F}\,,
\end{equation}
where $M_f^2 = m_1^2-m_2^2$ and $R$ is the worldsheet Ricci scalar. As we will see, when the fields are uncharged with respect to the monopole, the one-loop determinants only depend on the square masses $M_b^2$ and $M_f^2$. In case the fields do interact with the monopole, the fermionic spectrum is determined by not only $M_f^2$ but also by the sum and difference of $m_1$ and $m_2$, as we review in more detail in Appendix \ref{App: Fermionic monopole harmonics}. An important consistency condition on our operators, which ensures the cancellation of the conformal anomaly, are the sum rules \cite{Drukker:2000ep,Forini:2015mca,Cagnazzo:2017sny}
\begin{equation}\label{sumrules}
\begin{split}
\trace (-1)^F M^2 L^2 &= \trace  M_b^2L^2- \trace  M_f^2 L^2 =  R L^2 = 2\,,\\
\trace (-1)^F q^2 &= \trace q_b^2- \trace q_f^2 = 0\,, 
\end{split}
\end{equation}
where the trace runs over all modes in the theory.

Equipped with the eight bosonic and eight fermionic operators \eqref{operatordef}, the one-loop partition function of the non-interacting QFT living on the worldsheet can be written as
\begin{equation}
\text{Sdet}'{\mathbb{K}} = \f{\prod_{b}(\det' L^2{\cal K})}{\prod_{f}(\det' L{\cal D})}\,,
\end{equation}
where the product runs over all scalars and fermions in the theory and our conventions are summarised in appendix \ref{MeasureFactors}. The primes indicate that zero-modes should be excluded from the determinants. As mentioned above, we have rescaled the operators ${\cal K}$ and ${\cal D}$ before computing their determinant to remove the dependence of the scale $L$. This dependence is then reinstated at the level of $C(\chi)$ in the final answer. 

The spectral problem for our operators has been studied extensively in the past and we will simply borrow earlier results. Detailed analysis of the eigenfunctions and their properties can be found in \cite{Wu:1976ge,Wu:1977qk,Camporesi:1995fb,Borokhov:2002ib,Benini:2012ui,Pufu:2013vpa}. For our purpose we only need the eigenvalues of the operators and the corresponding degeneracies. For the bosonic operator $L^2{\cal K}$ the eigenvalues are
\begin{equation}
\lambda_b = l(l+1) - \f{q^2}{4} + M_b^2L^2\,,\qquad l -\f{|q|}{2}\in {\bf N}_0\,,
\end{equation}
and the corresponding degeneracies are $d_\lambda = 2l+1$. Therefore the logarithmic  of the functional determinant can be written as
\begin{equation}\label{bosonsum}
\log\det L^2{\cal K} = \sum_{j=(|q|+1)/2}^\infty(2j) \log\Big( j^2-\f14 - \f{q^2}{4} + M_b^2L^2 \Big)\,,
\end{equation}
where we have reindexed the sum $l= j-1/2$. For the fermions, we do not have eigenfunctions of the Dirac operator $L{\cal D}$ in the usual sense as the eigenvalues turn out to be `eigenmatrices' \cite{Pufu:2013vpa}. Nevertheless,  the logarithm of the fermionic functional determinant can be written down in a similar manner as for the bosons. The spectrum, however, depends on the charge of the fermions with respect to the monopole such that (see Appendix \ref{App: Fermionic monopole harmonics} for the details)\footnote{We thank Matteo Beccaria for pointing out an error in the first version.}
\begin{equation}\label{fermionsum}
		\log\det L{\cal D} = \sum_{l=|q|/2}^\infty (2l)\log\Big(l^2 - \frac{q^2}{4}  + M_f^2L^2\Big) - |q|\log L(m_1-\text{sgn}(q)m_2)\,,
\end{equation}
Note that when $q\rightarrow 0$ the spectrum indeed only depends on $M_f^2$, and not the individual values of $m_1$ and $m_2$, as mentioned above.

The infinite sums in \eqref{bosonsum} and \eqref{fermionsum} diverge and must be regularized. In order to treat both fermionic and bosonic sums simultaneously, we bring them in the form
\begin{equation}\label{spdef}
s_p(\mu) = \sum_{\substack{l = p\\l^2 \neq \mu}}^\infty{} (2l) \log\big( l^2-\mu \big)\,,
\end{equation}
by identifying the starting value $p$ of each sum
\begin{equation}
\begin{split}
p = (|q|+1)/2\,,&\qquad \text{for scalars}\,,\\
p = |q|/2\,,&\qquad\text{for fermions}\,,
\end{split}
\end{equation}
and defining the $\mu$ parameters as follows
\begin{equation}
\begin{split}
\mu = \f14 + \f{q^2}{4} - M^2L^2\,,&\qquad \text{for scalars}\,,\\
\mu = \f{q^2}{4} - M^2L^2\,,&\qquad\text{for fermions}\,. 
\end{split}
\end{equation}
The full logarithm of the partition function can therefore be expressed as a sum of functions $s_p(\mu)$ for some given values of $\mu$ and $p$ plus the fermionic contributions linear in $m_1$ and $m_2$. 
Before evaluating the sum, we remark that the argument of the $\log$ in \eqref{spdef} is positive. By hand we removed the zero-modes given by the states for which $l^2=\mu$ (with multiplicity $2l$), which have to be treated separately. 

As mentioned above, $s_p$ diverges, but a regularized version can be defined as follows
\begin{equation}
s_p^\text{reg}(\mu) = \sum_{\substack{l = p\\l^2 \neq \mu}}^\infty{} \Big[(2l) \log\big( l^2-\mu \big)-(4l) \log l+\f{2\mu}{l}\Big]\,,
\end{equation}
where the two last terms have been introduced to render the sum absolutely convergent. The middle term can be regularized by itself using the Hurwitz $\zeta$-functions
\begin{equation}
\sum_{l=p}^\infty (4l) \log l=  -4\zeta'(-1,p)\,,
\end{equation}
whereas the last term is divergent and it is formally given by
\begin{equation}\label{divergence}
\sum_{l=p}^\infty\f{2\mu}{l}= - 2\mu\,\psi(p)+2\mu\lim_{s\to 1}\f{1}{s-1} \,,
\end{equation}
where $\psi$ is the polygamma function. 
This is however not of immediate concern; when evaluating the complete string partition functions we will sum over all scalar and fermionic modes in the theory with relevant signs, and upon doing so the divergent pieces cancel. This cancellation is a non-trivial consequence of the sum rules \eqref{sumrules} satisfied by our modes, 
\begin{equation}
\trace (-1)^F \mu = 2  - \trace (-1)^F M^2L^2  + \f14\trace (-1)^F q^2 = 0\,.
\end{equation} 
With this in mind we can write
\begin{equation}
\bar s_p(\mu) = s_p^\text{reg}(\mu) - 4\zeta'(-1,p)+2\mu\,\psi(p)\,,
\end{equation}
where we dropped the divergent piece in \eqref{divergence} (indicated with the bar on $\bar s_p(\mu)$) which will cancel in string partition functions as explained.

The regulated sum $s_p^\text{reg}(\mu)$ can be evaluated by standard manipulations, in particular we note that $s_p^\text{reg}(0)= ({s}_p^\text{reg})'(0) =0$, where the prime denotes the derivative with respect to $\mu$. We can readily evaluate the second derivative of $s_p^\text{reg}$:
\begin{equation}
({s}_p^\text{reg})''(\mu) =\f{1}{2\sqrt{\mu}}\Big(\psi'(p+\sqrt{\mu})-\psi'(p-\sqrt{\mu})\Big)\,.
\end{equation}
Integrating once, and using that $({s}_p^\text{reg})'(0) =0$, we have
\begin{equation}\label{sprime}
({s}_p^\text{reg})'(\mu) =  \psi(p+\sqrt{\mu})+\psi(p-\sqrt{\mu}) - 2\psi(p)\,.
\end{equation}
To obtain a final value for $s_p^\text{reg}$ we integrate \eqref{sprime} once again to obtain,
\begin{equation}
{s}_p^\text{reg}(\mu) =  -2\mu\psi(p)+\int_0^\mu\Big(\psi(p+\sqrt{x})+\psi(p-\sqrt{x})\Big)\dd x \,.
\end{equation}
With this we end with our final result which we will use below to evaluate our string partition functions
\begin{equation}
\bar{s}_p(\mu) =  - 4\zeta'(-1,p)+\int_0^\mu\Big(\psi(p+\sqrt{x})+\psi(p-\sqrt{x})\Big)\dd x \,.
\end{equation}
This is the expression we will utilize when we evaluate the string partition function in the following sections.

\subsection{Higher instantons and orbifold partition function}
\label{ssec:Higherinsta}

As previously discussed, in the saddle point expansion of the string partition function, we must sum over all non-trivial saddles of the string and quantize its fluctuations around the classical solution. The resulting series contains infinitely many terms since we can always wrap the string multiple times around the same cycle in the target manifold. The classical action for such a multiply wound string is simply $n$ times the classical action of a single string, where $n$ is the winding number of the string around the cycle in question. In this case, we can view the worldsheet as a sphere, where the north and south poles have an angular excess of $2\pi(n-1)$. 
In the fermionic and bosonic operators controlling the one-loop fluctuations of the multi-wound string, the same masses appear as in the case $n=1$. 
However, the partition function of the multi-wound string is different because the spectrum of the operators depends on the underlying two-dimensional manifold. 
We propose that the one-loop partition function for a multi-wound string is given by the one-loop partition function for the single string times contributions attributable to the poles of the sphere. 
The intuition here is that we can recover the single string partition function by cutting out the poles of the sphere and replacing them with a smooth cap. 
However, at each pole, this replacement yields an error given by the difference between a partition function for a smooth disk and one for a disk with a conical singularity (in particular here with an angular excess).  
Our conjecture is therefore that the multi-wound sphere partition function is given by
\be
Z^{(n)} = \e^{-n S_\text{cl}} Z_\text{1-loop}\, z_n^2\,,
\ee
where $Z_\text{1-loop}$ is the single string one-loop partition function in Eq. \eqref{eq:1-loop pf def} and $z_n$ is the local factor received from each pole, as explained.

In order to determine the factor $z_n$, we revisit the computation of the disk partition function in AdS$_5\times S^5$ dual to the 1/2-BPS Wilson loop in ${\cal N}=4$ SYM. The multi-wound string computation has been studied extensively in \cite{Kruczenski:2008zk,Buchbinder:2014nia,Bergamin:2015vxa,Forini:2017whz}, but a match with the answer obtained by supersymmetric localization has not yet been obtained. We will not be able to give a first principle derivation of the partition function of the multi-wound string, but instead we will use the exact evaluation of the Wilson loop expectation value from the field theory \cite{Pestun:2007rz} to fix the value of $z_n$.

Recall that the 1/2-BPS Wilson loop in ${\cal N}=4$ SYM is dual to an open string that attaches to the boundary of AdS$_5$. If we use spherical coordinates on AdS$_5$
\be
\dd s_{\text{AdS}_5}^2 = \dd \rho^2 + \sinh^2\rho\,\dd\Omega_4^2\,,
\ee
the string in question sits on the equator of the four-sphere and wraps the $\rho$ direction. On $S^5$ the string sits at any fixed location which can be taken to be the north pole of $S^5$. The string partition function for the single string has been analyzed in \cite{Drukker:2000ep,Kruczenski:2008zk,Kristjansen:2012nz} and nicely summarized recently in \cite{Giombi:2020mhz}. The classical action of the string is simply the area of AdS$_2$ which, after regularization, takes the value ${\cal A} = -\sqrt{\lambda}$ where $\lambda$ is the ${\cal N}=4$ 't~Hooft coupling constant related to the radius of AdS$_5$ by $L^2 = \ell_s^2\sqrt{\lambda}$. The string coupling constant $g_s$ is related to $\lambda$ and the rank of the gauge group $N$ via
\be
g_s = \f{\lambda}{4\pi N}\,,
\ee
and the Fradkin-Tseytlin contribution to the string partition function is $g_s^{-1}$. The one-loop partition function of the string modes themselves is given by \cite{Kruczenski:2008zk,Buchbinder:2014nia}\footnote{There are no zero-modes in this case.}
\be
(\text{Sdet}{\mathbb{K}})^{-1/2} = \f{1}{\sqrt{2\pi}}\,,
\ee
and so the single-string partition function can now be assembled according to \eqref{saddlepointexpansion} and \eqref{eq:1-loop pf def} using eq. \eqref{C1eq}
\be
Z_\text{string} = \f{ N}{\lambda^{3/4}}\sqrt{\f{2}{\pi}}\e^{\sqrt{\lambda}}\,,
\ee
which matches the prediction of the QFT for the Wilson loop expectation value~\cite{Erickson:2000af,Drukker:2000rr,Pestun:2007rz}. 
Now we consider the $n$-wound string, dual to the $n$ wound Wilson loop operator whose vacuum expectation value is \cite{Drukker:2005cu,Pestun:2007rz} 
\be\label{multiWLanswer}
\langle {\cal W}\rangle = \f{ N}{n^{3/2}\lambda^{3/4}}\sqrt{\f{2}{\pi}}\e^{n\sqrt{\lambda}}\,.
\ee
Following our previous logic for the sphere partition function, now we should be able to obtain the multi-wound partition function by modifying only the classical action and adding the multiplicative factor $z_n$. In this case we only have one power of $z_n$ since there is only one orbifold point on the worldsheet which must be counted. Comparing the field theory prediction \eqref{multiWLanswer} with our conjecture, we conclude that\footnote{It would be interesting to confirm a similar behavior for Wilson loops in the ABJM theory. 
To this end it is important to identify the correct representation of the relevant Wilson loop in the $\U(N)_k \times \U(N)_{-k}$ gauge group that is dual to the multi-wound fundamental string.
}
\be\label{znformula}
z_n = \f{1}{n^{3/2}}\,.
\ee
This means that for spherical worldsheets, the partition function for the multi-wound string is divided by a factor of $n^3$ as compared to the single string (modulo the classical action). It is interesting to note that such a contribution has indeed been observed for the topological string where a contribution of a multi-wound rational curve in a Calabi-Yau threefold is assigned exactly a cube factor \cite{Aspinwall:1991ce}. 

We conclude by noting that for a multi-wound string we encounter orbifold points on the worldsheet with conical excess angles. In other situations we may encounter orbifold points with a conical deficit. In this case we expect that the only difference is the appearance of the factor $n^{3/2}$ in the numerator rather than in the denominator as in \eqref{znformula}.

 \section{Instanton corrections to 5D Free energy}
 \label{sec:5dsym}
 
In this section we will discuss non-perturbative terms in the $S^5$ partition function of maximally supersymmetric Yang-Mills theory in five dimensions. At large $N$ the partition function takes a particularly simple form (see \eqref{5DLogZ}) and as we will see, all non-perturbative terms in the expression are due to worldsheet instantons. We will also discuss the relation to the giant graviton expansion of the six-dimensional (2,0) theory.
\subsection{Field theory}	
\label{ssec:5dsym FT}
When placed on a round five-sphere, Euclidean supersymmetric Yang-Mills theory in five dimensions can be localized to a matrix model \cite{Kim:2012ava}. Remarkably the matrix model encountered is exactly the same as the one for Chern-Simons (CS) theory in three dimensions which has been studied extensively in the past, notably in \cite{Witten:1988hf}. In the context of five-dimensional SYM, the matrix model depends on the two microscopic parameters of the underlying theory, i.e. the rank of the gauge group $N$ (we work with gauge group $\U(N)$) and a dimensionless combination of the Yang-Mills coupling constant $g_\text{YM}^2$ and the five-sphere radius ${\cal R}$. We find it useful to express our formulae in terms of a dimensionless 't~Hooft coupling constant $\xi$ which is defined as 
\begin{equation}\label{Eq: 't Hooft coupling in 5d}
\xi =  \frac{g_\text{YM}^2 N}{2\pi \mathcal R}\,.
\end{equation}
In what follows we will study the partition function of the five-dimensional theory in the large $N$ limit while keeping $\xi$ finite. As we will recall in greater detail below, five-dimensional SYM at strong coupling is related to the six-dimensional (2,0) theory where the size of the sixth dimension is related to the coupling constant of the 5D theory. The five-dimensional supersymmetric partition function on $S^5$ equals an unrefined superconformal index of the six-dimensional theory on $S^5\times S^1_\beta$, where the periodicity of the $\beta$-cycle is directly related to the gauge coupling through $\xi = N\beta$ \cite{Kim:2012ava,Kallen:2012zn,Kim:2012tr,Lockhart:2012vp,Kim:2012qf,Minahan:2013jwa,Kim:2013nva}. When expanding the five-dimensional  partition function around small $\rme^{-\beta}$ one finds integer coefficients, giving evidence that the $S^5$ partition function can be viewed as the index of the (2,0) theory. In this paper we will study the partition function in an expansion around large $N$ and $\xi$, effectively taking a Cardy-like limit of the six-dimensional index. 
	
On the holographic side, the six-dimensional $(2,0)$ theory is described by M-theory on AdS$_7 \times S^4$, and the high temperature limit corresponds to a circle reduction to type IIA string theory. Using this IIA limit, in section \ref{ssec:5dsym ST}, we will be able to compute exponentially suppressed contributions to the partition function coming from worldsheet instantons in the internal geometry. To corroborate our results for the string theory instanton contributions to the partition function we will compare to known field theory results, which we will now summarize. 
	
The $S^5$ free energy $F=-\log Z$ of SYM computed by supersymmetric localization is a sum of the perturbative partition function $F_{\text{pert}}$ and an instanton contribution collected in $F_{\text{inst}}$. Explicitly these terms take the form \cite{Kim:2012ava}
\begin{equation}
\begin{aligned}
F_{\text{pert}} =&\, -\log\bigg[\left( \frac{\rmi \beta}{2\pi} \right)^{N/2} \rme^{\frac{N(N^2-1)}{6} \beta} \prod\limits_{m=1}^{N-1} (1- \rme^{-\beta m})^{N-m}\bigg]\,,\\
F_{\text{inst}} =&\, 
N\log{\eta\Big(\frac{2\pi\rmi}{\beta}\Big)}\,,
\end{aligned}
\end{equation}
where $\eta(z)$ is the Dedekind-$\eta$ function. After substituting $\beta=\xi/N$ and taking the large $N$ limit we find that the perturbative free energy reduces to%
\footnote{Taking the large $N$ limit of the expressions (3.42) in both \cite{Kim:2012ava} and \cite{Kim:2012qf} is  subtle and one has to first use the modularity property of the Dedekind-$\eta$ function stating that $\eta(-1/z) = \sqrt{-\rmi z}\, \eta(z)$. This result can also be deduced from the saddle point analysis of the five-dimensional matrix model which admits an analytic solution. We refer to \cite{Marino:2004eq} where the equivalent matrix model for three-dimensional CS theory was solved in this limit.} 
\begin{equation}\label{5DLogZ}
F_{\text{pert}} 
=  -N^2 \bigg( \frac{\xi}{6} - \frac{\pi^2}{6\xi} + \frac{\zeta(3)}{\xi^2} - \frac{\text{Li}_3(\rme^{-\xi})}{\xi^2} \bigg) + \mathcal O(N\log N)\,,
\end{equation}
and contributes at order $N^2$ as expected for a gauge theory. On the other hand, we can explicitly verify that instantons do not contribute at leading order
	since the Dedekind-$\eta$ function has an exponential suppression at large $N$, allowing us to disregard it at leading order $N^2$.
At weak coupling the free energy can be expanded 
\be
{F_{\text{pert}} \over N^2}=
\frac{3}{4}
-\frac12 \log\xi
-\frac{\xi}{12}
-\frac{\xi^2}{288}
+ \frac{\xi^4}{86400}
+  {\cal O}(\xi^6)\,,
\ee
and contains infinitely many terms in its series. 
On the other hand, the full expression  \eqref{5DLogZ} is exact in $\xi$ and its strong coupling expansion is finite, containing only three terms.
This is a remarkable property of the five-dimensional partition function and is most easily explained by its close relation to the index of the six-dimensional theory. In addition to the strong coupling perturbative series being simple, the non-perturbative terms in $\xi$ 
\begin{equation}\label{TheLiterms}
\frac{N^2}{\xi^2}\text{Li}_3(\rme^{-\xi}) = \frac{N^2}{\xi^2}\sum_{n=1}^\infty \frac{\rme^{-n \xi}}{n^3} \,,
\end{equation}
	are also noteworthy in that they are all of the same type and, as we will now explicitly verify, represent worldsheet instantons on the string side.

\subsection{String theory}
\label{ssec:5dsym ST}
The holographic dual to five-dimensional SYM on $S^5$ was constructed in \cite{Bobev:2018ugk} and describes the back-reaction of spherical D4 branes in ten-dimensional type IIA supergravity. This background is most easily recovered by dimensional reduction of the AdS$_7\times S^4$ solution of eleven-dimensional supergravity. In order to recover the spherical brane background, one must write AdS$_7$ in global coordinates, analytically continue the time direction which is then reduced over. Direct dimensional reduction over Euclidean time breaks supersymmetry but 16 real supercharges can be preserved by a reduction over the diagonal combination of the Euclidean time and an $S^1$ inside $S^4$. We refer to  \cite{Bobev:2018ugk,Gautason:2021vfc} for more details. Since we will work exclusively in the type IIA limit (given that we are interested in quantizing strings in the ten-dimensional background) we will only write the ten-dimensional reduction. The metric is given by
	\begin{equation}\label{5Ddualmetric}
		\rmd s^2 = \f{L^2}{\sqrt{Q}}\,\Big[4\big(\dd\rho^2+ \sinh^2\rho\,\dd\Omega_5^2\big)+ \dd\theta^2 + \cos^2\theta\, \dd \Omega_2^2  + Q \cosh^2\rho\sin^2\theta\, \dd \phi^2\Big]\,,
	\end{equation}
	where $\rho$ plays the role of a holographic direction and for large $\rho$ we recover the metric for the near-horizon region of D4 branes in flat space. The angular variables $\theta$ and $\phi$ together with the two-sphere $\dd\Omega_2^2$ form a topological four-sphere which is squashed by the presence of the function $Q$ which is
	\begin{equation}
		Q = \f{4}{4\cosh^2\rho- \sin^2\theta}\,.
	\end{equation}
	The background length scale can be mapped to the field theory 't Hooft coupling using the string length \cite{Bobev:2019bvq}
	\begin{equation}
		L^2 = \ell_s^2 \xi\,,
	\end{equation}
	and strong coupling $\xi\gg 1$ implies that $L\gg \ell_s$, such that the string saddle point approximation is valid.
	The dilaton and form fields are given by
	\begin{equation}
		\begin{aligned}
			\e^{2\Phi} &= \f{\xi^3}{N^2\pi^2}\f{1}{Q^{3/2}}\,,\\
			B_2 &= i\f{ L^2}{2}\cos^3\theta\,\vol_{S^2}\,,
		\end{aligned}
	\qquad 
	\begin{aligned}
		C_1 &= i\f{\pi N L}{2\xi^{3/2}}Q\sin^2\theta\,\dd\phi\,, \\
		C_3 &= \frac{\pi N L^3}{\xi^{3/2}} \cos^3\theta\,\dd\phi\wedge\vol_{S^2}\,.
	\end{aligned}
	\end{equation}
Using this solution, the leading order free energy can be computed by evaluating the on-shell supergravity action. This is done by considering the dimensional reduction of type IIA to six-dimensional supergravity where the solution can be readily embedded. After cancelling divergences and adjusting finite counterterms the on-shell action is found \cite{Bobev:2019bvq}
\begin{equation}
S_\text{sugra} = -N^2 \frac{\xi}{6}\,,
\end{equation}
which matches the leading order term in the  QFT answer \eqref{5DLogZ}. It would be interesting to extend this match to higher order by evaluating the one-shell action of type IIA supergravity including higher derivative terms. In similar spirit, the vacuum expectation value of 1/2-BPS Wilson loop was reproduced by evaluating a string theory on-shell action  in this background\cite{Bobev:2019bvq}. In \cite{Gautason:2021vfc} this match was extended to next-to-leading order by quantizing the fluctuations of the string and computing its one-loop partition function.

We are interested in supersymmetric configurations for a spherical string in this geometry with finite classical action. This will give rise to a non-perturbative correction to the free energy as explained in section \ref{sec:GS}. Since there are no two-cycles in our geometry, one is tempted to conclude that there are no stable configurations for such a string. However, due to the presence of an NSNS $B_2$ field, there is a supersymmetric configuration for the string which is supported by the 2-form. This supersymmetric embedding is for a string that wraps the $S^2$ threaded by the $B_2$ field, and sits at%
\footnote{Supersymmetry of the embedding can be explicitly verified using the calibration form in \cite{Mezei:2018url}.}
\be
\rho=  0 = \theta\,.
\ee
At this point, the remaining $S^5$ and $S^1$ collapse to zero size and the worldsheet metric takes the form of a round sphere as \eqref{sphericalworldsheet} with $L^2=\ell_s^2 \xi$. The solution therefore preserves $\SO(3)\times\SO(6)\times\SO(2)$ symmetry where the first $\SO(3)$ represents the rotation symmetry of the worldsheet itself.
	
The classical action \eqref{classicalAction} is now easily evaluated for our string configuration. In particular we find
\begin{equation}\label{5d-classical-action}
\mathcal A = 2\xi\,,\quad \text{and} \quad \frac{\rmi}{2\pi \ell_s^2}
\int B_2 = -\xi\,,
\end{equation}
such that $S_{\text{cl}} = \xi$. 	Wrapping $n$ fundamental strings around this calibrated cycle, the classical action is simply multiplied by the total number of strings and we find $S_\text{cl} = n\xi$. This shows that using the holographic relation \eqref{Eq:Holography}, the classical action correctly reproduces the exponentially behaviour of the subleading terms \eqref{TheLiterms} in the sphere free energy.

For the one-loop partition function we will now focus on the single instanton case. Higher rank instantons will be discussed in section \ref{Sec: higher instanton 5d SYM}. As explained in section \ref{sec:GS}, we need to compute the FT term and the determinant of the kinetic operators for the modes on the string. Since the dilaton is constant on the worldsheet, the former is rather simple for our solution and equals
\begin{equation}
\rme^{-S_{\text{FT}}}=\frac{N^2 \pi^2}{\xi^3}\,.
\end{equation}
This correctly reproduces the $N^2$ behaviour of the prefactor in \eqref{TheLiterms} as expected from considering spherical worldsheets. For the one-loop determinant we follow the procedure outlined in section \ref{ssec:oneloop}. Expanding the Green-Schwarz action as discussed in section \ref{ssec:oneloop} we obtain operators of the form \eqref{operatordef} for the eight fermions and eight bosonic modes. Since the worldsheet is embedded trivially in the ambient geometry we do not encounter any monopole field meaning that all 16 modes are uncharged, $q=0$. Explicitly we find the masses and degeneracies as listed in table \ref{5Dspectrum table}.
\begin{table}[ht]
\begin{center}
{\renewcommand{\arraystretch}{1.1} 
\begin{tabular}{@{\extracolsep{10 pt}}l l c c c c}
\toprule
Field&Degeneracy&  $M^2L^2$ & $q$&  $\mu$ &  $p$\\
\midrule
\noalign{\smallskip}
scalars& 6&$\frac14$&0&0&$\frac12$\\
&2&$-\frac34$&0&1&$\frac12$\\
\midrule
fermions&8&$-\frac14$&0&$\frac14$&1\\
\bottomrule
\end{tabular}}
\caption{\label{5Dspectrum table}The spectrum of fluctuations of string modes around the classical instanton solution in the geometry \eqref{5Ddualmetric}. The quantities $\mu$ and $p$ are defined in section \ref{ssec:oneloop}.}
\end{center}
\end{table}
	Using this spectrum, we can use the general discussion in section \ref{sec:GS} to compute the regularized one-loop partition function
	\begin{equation}\label{logminus4}
	-\f12 \log ({\text{Sdet}' \mathbb{K}}) = 4s_1(1/4) - 3 s_{1/2}(0) - s_{1/2}(1) = \log(-4)\,,
	\end{equation}
	and we conclude that
	\begin{equation}
	({\text{Sdet}' \mathbb{K}})^{-1/2} = -4\,.
	\end{equation}
	Putting everything together we find that the single instanton partition function is given by
	\begin{equation}
		Z_{\text{inst}}^{(1)} =- C(2)\frac{4N^2 \pi^2}{\xi^3}\rme^{-\xi} \,.
	\end{equation}
	As mentioned in section \ref{sec:GS} the only remaining task is to explicitly determine $C(2)$, which we can do by comparing to the field theory result, out of which we find that
	\begin{equation}
		C(2) = \frac{\mathcal A}{8\pi^2}\,.
	\end{equation}
	By using \eqref{5d-classical-action}, this gives for the string partition function
	\be
	Z_{\text{inst}}^{(1)} =-\frac{N^2}{\xi^2}\rme^{-\xi} \,.
	\ee
	Note that in $C(2)$ there is a factor $-2$ difference when comparing to $C(1)^2$. Equipped with its value \eqref{C2eq}, we can use this procedure to compute sphere partition functions in much the same way in other holographic geometries. 
	\subsection{Higher instantons}\label{Sec: higher instanton 5d SYM}
	Let us now discuss higher instantons. These are represented by a stack of multiple strings that wrap the same 2-sphere supported by $B_2$. The fact that this is the only BPS configuration at each higher instanton level is clear in the IIA limit. However, from an eleven-dimensional perspective this may not be obvious since presumably there are two different cycles that M2  branes can wrap in $S^4$. This point will be discussed in more detail in section \ref{ssec:giantgravitons}. To determine the contributions of higher order instantons we have to compute the string partition function on the 2-sphere with angular excess at both poles. As we have argued in section \ref{ssec:Higherinsta}, the partition function of the multi-wound string is a product of the single string partition function times the contribution from the two orbifold points. In other words, we have
	\begin{equation}
		Z_{\text{inst}}^{(n)} = C(2)({\text{Sdet}' \mathbb{K}})^{-1/2} z_n^2\text{e}^{-S_{\text{FT}}}\text{e}^{-n\xi} = - \frac{N^2 }{\xi^2 n^3} \text{e}^{-n\xi} \,.
	\end{equation}
	These contributions can be summed to correctly match the field theory answer \eqref{TheLiterms}, yielding 
	\begin{equation}\label{Eq: arbitrary higher instanton for 5d SYM}
		\sum\limits_{n=1}^{\infty} Z_{\text{inst}}^{(n)} = - \frac{N^2}{\xi^2} \text{Li}_3(\text{e}^{-\xi})\,.
	\end{equation}
	Notice in particular that the sign of the instanton contribution is correctly reproduced by our computation where, on the string theory side, the minus originates from the one-loop partition function \eqref{logminus4}.
	
	In this section we have utilized the precise knowledge of the QFT partition function to fix the coefficient $C(2)$ and to test our conjecture for higher instantons, leading us to a complete holographic match with the non-perturbative instanton contributions to the field theory partition function in the large $N$ limit. In the sections below we will utilize these results to compute worldsheet instanton contributions in type IIA backgrounds dual to three-dimensional CS theories. Before doing so, we will relate our results of this section to the giant graviton expansion of M2-branes in eleven dimensions.	
	\subsection{Giant gravitons in M-theory}\label{ssec:giantgravitons}
	As discussed above, the type IIA solution of interest can be uplifted to eleven dimensions, for the details of this uplift we refer to \cite{Bobev:2018ugk,Gautason:2021vfc}. The eleven-dimensional solution is topologically AdS$_7 \times S^4$, with an $S^5 \times S^1_\beta$ boundary in the AdS space, on which the dual six-dimensional $(2,0)$ theory resides. This M-theory uplift clarifies the connection between the $S^5$ partition function of maximal SYM, and the superconformal index of the $(2,0)$ theory discussed in section \ref{ssec:5dsym FT}. As described in \cite{Kim:2012ava}, the $S^5$ partition function only captures an unrefined superconformal index of the $(2,0)$ theory. This will have a direct consequence on the dual M-theory (and string theory) observables that we will discuss in this section. 
	
	In \cite{Arai:2020uwd} it was argued that the superconformal index of the $(2,0)$ theory with gauge group $\U(N)$ can be reproduced in the M-theory description by the Kaluza-Klein index times finite $N$-corrections coming from supersymmetric M2-brane excitations in eleven dimensions, also known as giant gravitons \cite{McGreevy:2000cw}.\footnote{A similar claim was initially made for the superconformal index of $\mathcal N=4$ SYM in four dimensions \cite{Biswas:2006tj,Gaiotto:2021xce}, and for the related S-fold theories in \cite{Arai:2019xmp}.}  Explicitly, the superconformal index was conjectured to take the form
	\begin{equation}\label{Eq: Imamura and co conjecture for finite N index in 2,0 theory}
		\mathcal I^{\text{U}(N)} = \mathcal I^{\text{U}(\infty)} \left(1 + \sum\limits_C \mathcal I_C^{\text{M2}}\right)\,, 
	\end{equation}
	where $\mathcal I^{\text{U}(\infty)}$ is the index in the strict supergravity limit, determined by the Kaluza-Klein modes in the M-theory background, and $\mathcal I_C^{\text{M2}}$ are finite $N$ corrections due to giant gravitons for which the sum runs over all possible supersymmetric configurations of the M2-branes. There are two distinct supersymmetric M2-brane embeddings wrapping $S^2 \times S^1_\beta$, where the $S^2$ resides in $S^4$ \cite{Mikhailov:2000ya}. These corrections were argued to be computed by localizing the ABJM theory living on the supersymmetric M2-branes. We are interested in a Schur-like limit of this index, where the M2-branes are only allowed to wrap one of the two-cycles in $S^4$, producing an unrefined index as in the field theory analysis of \cite{Kim:2012ava}. Upon a reduction to type IIA string theory, described in section \ref{ssec:5dsym ST}, these M2-brane giant gravitons reduce to the worldsheet instantons computed in equation \eqref{Eq: arbitrary higher instanton for 5d SYM}. 
	The giant graviton expansion has been computed up to finite order of M2-branes in \cite{Imamura:2022aua}. It will be very interesting to extend this computation to arbitrary order of the number of giant gravitons, in the unrefined limit, to reproduce and extend our result in section \ref{Sec: higher instanton 5d SYM}. 
	
	Finally, we would like to note that the M2-brane picture, employing supersymmetric localization, also provides an explanation for the simple and exact functional form of our answer in equation \eqref{Eq: arbitrary higher instanton for 5d SYM}, coming from a one-loop computation of the worldsheet instanton.

\section{Instanton corrections to the ABJ(M) \texorpdfstring{$S^3$}{S3} free energy}
\label{sec:ABJM}
The second example  we will study is the AdS$_4 \times \CP^3$ background in type IIA string theory. Its holographic dual is the ABJ(M) theory, at large $N$ and large CS level $k$. We will keep our discussion general to include also the ABJ theory in addition to ABJM. This difference is usually not visible in the supergravity limit, but it is visible to worldsheet instantons as we will see. In the following section we will briefly review the ABJ(M) theories, focusing on known results for its $S^3$ supersymmetric partition function, both from the QFT and string theory sides. In section \ref{ssec:ABJM ST} we will turn to the string theory side and compute worldsheet instanton contributions to the string partition function in the AdS$_4 \times \CP^3$ background and show that the string theory computation matches known results from the QFT. Finally, in section \ref{ssec:higher instantons ABJM} we discuss the higher level instantons from the QFT and the string theory perspective.
\subsection{Field theory and known holographic results}
\label{ssec:ABJM FT}
The ABJM theory \cite{Aharony:2008ug} is a three-dimensional ${\cal N}=6$ superconformal field theory with two gauge groups and opposite CS level $\U(N)_k \times \U(N)_{-k}$. In ${\cal N}=2$ language the theory consists of four chiral multiplets $(X_{i=1,2}, Y_{i=1,2})$ that transform in the bifundamental representation of the gauge nodes and have $\U(1)_R$-charges which are all equal to $R[X_i] = R[Y_i] = 1/2$. In section \ref{sec:3dCS} we will also discuss more general R-symmetry assignment which is consistent with supersymmetry. These can be viewed as mass deformations of the ABJM theory.

In \cite{Aharony:2008gk} a different extension of the ABJM theory was considered which is called the ABJ theory after the original authors. In this case the CS level of both gauge nodes are equal but their rank is not $\U(N+l)_k \times \U(N)_{-k}$. This extension will be helpful for our analysis of worldsheet instantons in the dual geometry and so we will adjust our discussion to include also this case.

The ABJ(M) $S^3$ partition function  was originally localized in \cite{Kapustin:2009kz}, reducing the path integral to a matrix model. To solve this matrix model analytically is still a daunting task. However, it was found in \cite{Marino:2009jd} that at large rank the matrix model is dual to the  topological string
on $\mathbb{F}_0 = \mathbb{P}^1 \times \mathbb{P}^1$, 
which can be solved in an $1/N$ expansion. Further analyses of the matrix model were carried out in \cite{Drukker:2010nc,Drukker:2011zy} which ultimately led \cite{Fuji:2011km,Marino:2011eh} (and later \cite{Matsumoto:2013nya,Honda:2014npa} in the case of ABJ) to evaluate the topological string perturbation series into an Airy function of the form
\begin{equation}\label{eq:ABJM free energy}
	F_{\text{pert}} =  -\log\bigg[C^{-1/3} \rme^{A} \text{Ai} \Big( C^{-1/3} (N - B) \Big)\bigg]\,,
\end{equation}
where
\begin{equation}\label{Eq: defs of ABC in Airy}
	 A = C \zeta(3) \left(1-\frac{k^3}{16}\right) - \frac{k^2}{\pi^2} \int\limits_0^\infty \rmd x \frac{x\log(1-\rme^{-2x})}{1- \rme^{k x}}\,,\quad B = \frac{k}{24} + \frac{1}{3k}+\f{l^2}{2k}-\f{l}{2} \,, \quad C = \frac{2}{\pi^2 k}.
\end{equation}

The holographic dual to the ABJ(M) theory is provided by the AdS$_4\times S^7/\Z_k$ background of eleven-dimensional supergravity \cite{Aharony:2008ug,Aharony:2008gk}. 
The freely acting $\Z_k$ orbifolding acts on the Hopf fiber of $S^7$ and in the large $k$ limit the holographic dual is more accurately given in terms of the AdS$_4\times \CP^3$ solution of type IIA supergravity. The difference in the rank of the gauge groups, $l$, is translated in the holographic dual to a $B_2$ period on $\mathbf{C}P^1\subset \mathbf{C}P^3$ \cite{Aharony:2008gk,Aharony:2009fc}. Reproducing the full perturbative answer \eqref{eq:ABJM free energy} with a purely gravity computation is still out of reach. The reason is that the subleading corrections in the large $N$-expansion  arise from higher derivative corrections to the type IIA/11d supergravity action, which are currently unknown. In \cite{Dabholkar:2014wpa} a different approach was taken, namely supersymmetric localization in the supergravity bulk theory was employed to argue for the Airy function behavior of the partition function. More recently, in \cite{Bobev:2020egg,Bobev:2021oku} the problem was reduced from eleven dimensions to a four-dimensional one, in which the leading higher derivative corrections to the action could be constrained by the structure of the four-dimensional theory to a number of unknown constants which were then fixed by comparison to the QFT. This resulted in a holographic match of the free energy in the large $N$ expansion to order $N^{1/2}$.

The perturbative answer in \eqref{eq:ABJM free energy} are known to have non-perturbative corrections, which are the main focus of our work. The study of these non-perturbative corrections was already initiated in the original papers \cite{Drukker:2010nc,Drukker:2011zy}. Shortly after, in a series of papers \cite{Marino:2011eh,Hatsuda:2012dt,Hatsuda:2013gj,Hatsuda:2013oxa,Calvo:2012du,Hatsuda:2012hm,Putrov:2012zi}, the analysis of instanton corrections was extended by using the Fermi gas formulation of the matrix model resulting in a conjectured exact answer as function of $\lambda=N/k$ and $N$, for every instanton level. Generically, these instanton contributions to the free energy take the form
\begin{equation}\label{eq:inst in ABJM}
	F_{\text{inst}} = \sum\limits_{m,n} c_{m,n}(N,l,\lambda) \rme^{-\sqrt{2}\pi (n N/\sqrt{\lambda} + 2 m \sqrt{\lambda})}\,,
\end{equation}
where $(m, n)$ are non-negative integers with $(m,n)\neq (0,0)$. 
Using the $\mathbb{F}_0$ 
topological string picture, the coefficients $c_{m,0}$ and $c_{0,n}$ could be systematically computed. The generic coefficients $c_{m,n}$ were predicted to be completely fixed by $c_{m,0}$ and $c_{0,n}$ \cite{Hatsuda:2013gj}.

The two integers $n$ and $m$ in \eqref{eq:inst in ABJM} enumerate two different instanton effects. On the type IIA string theory side, these count D2 branes wrapping $\RP^3 \subset \CP^3$  and fundamental strings wrapping $\CP^1 \subset \CP^3$ respectively. In the eleven-dimensional description, the D2 brane and the strings both uplift to M2 branes wrapping different 3-cycles in $S^7/\Z_k$. We are interested in the type IIA limit, $N\gg 1$ and $\lambda \gg 1$, in which  the worldsheet instantons coming from strings wrapping $\CP^1\subset \CP^3$ dominate over the D2 brane instantons. Our aim will be to use string theory directly to compute the leading worldsheet instanton corrections to the ABJ(M) free energy. Specifically, in this limit, the first non-perturbative worldsheet instanton correction to the free energy is \cite{Drukker:2010nc}\footnote{There is a minus sign discrepancy with \cite{Drukker:2010nc} where the convention is used that $F_\text{there} = + \log Z$.}
\begin{equation}\label{eq:field theory instanton ABJM}
 \frac{N^2}{(2\pi \lambda)^2}  \cos \Big(\f{2\pi l}{k}\Big)\,  \rme^{-2 \pi \sqrt{2 \lambda} }
\end{equation}
Below, in section \ref{ssec:ABJM ST}, we will apply the techniques described in section \ref{sec:GS} to reproduce this expression 
directly from string theory by computing the one-loop partition function of strings wrapping $\CP^1 \subset \CP^3$. In section \ref{ssec:higher instantons ABJM} we discuss the higher rank instantons briefly reviewing the field theory prediction in \cite{Drukker:2010nc,Hatsuda:2012dt,Hatsuda:2013gj,Hatsuda:2013oxa}. We will then use string theory to partly recover these results.

\subsection{String theory}
\label{ssec:ABJM ST}

We start by summarizing the type IIA supergravity background dual to the ABJ(M) theory on $S^3$. The supergravity background metric is 
\begin{equation}\label{Eq: ABJM background metric}
	\rmd s^2 = L^2 \left( \rmd s_{\text{AdS}_4}^2 + 4 \rmd s_{\CP^3}^2 \right)\,,
\end{equation}
where AdS$_4$ has an $S^3$ boundary such that
\begin{equation}\label{AdS4metric}
	\rmd s^2_{\text{AdS}_4} = \rmd \rho^2 + \sinh^2 \rho\, \rmd \Omega_3^2\,,
\end{equation}
with $\rmd \Omega_3^2$ the round metric on $S^3$. The string length scales and coupling are related to the coupling constant $\lambda$ and the rank of the gauge group through 
\begin{equation}
	L^2 = \pi\ell_s^2\sqrt{2\lambda}\,,\quad \rme^{2\Phi}=g_s^2=\frac{\pi(2\lambda)^{5/2}}{N^2} \,.
\end{equation}
The form fields take the form
\begin{equation}
\begin{split}
C_1 &= \frac{2N\ell_s}{{\lambda}} \, \omega\,,\\
B_2 &= \left(\f12 -\frac{ l }{k}\right)4\pi\ell_s^2J\,,\\
C_3 &= -\frac{3 i \pi N \ell_s^3}{\sqrt{2\lambda}} \, U \vol_{S^3} \,,
\end{split}
\end{equation}
where $\vol_{S^3}$ is the volume form on the (unit radius) $S^3$ and 
\be\label{Uundeformed}
U = \f{1}{12}(-9\cosh\rho + \cosh 3\rho)\,,
\ee
such that $\partial_\rho U = \sinh^3\rho$ and we get the proper volume form on AdS$_4$ appearing in $\dd C_3$. In the above form fields we also introduced the K\"ahler form $J=\dd\omega$ on $\CP^3$ and its integral $\omega$. We have normalized our K\"ahler form in the standard way, such that the volume form on the unit radius $\CP^3$ is given by $J^3/6$ and the volume itself is $\pi^3/6$. Notice that we have added a period to the $B_2$ gauge potential along $J$. The three-form $H_3 = \dd B_2$ vanishes for this gauge potential but it plays a role when distinguishing the ABJM theory with two equal gauge groups $\U(N)_k\times \U(N)_{-k}$ from the ABJ theory with different gauge groups $\U(N+l)_k\times \U(N)_{-k}$  where $0\le l < k$\cite{Aharony:2008gk}.
The rank mismatch $l$ can be observed from the quantized D4 brane charge\footnote{The shift with $k/2$ was explained in \cite{Aharony:2009fc} to be a direct consequence of quantization conditions in the background.}
\be
\f{1}{(2\pi \ell_s)^3} \int_{\CP^2} (F_4+H\w C_1) = l - \frac{k}{2} \,.
\ee
Even though the difference between the different ABJ theories with $l=1,2,\cdots$ and the ABJM theory with $l=0$ is subleading for most observables in the IIA limit, the difference is visible to worldsheet instantons as was originally noted in \cite{Aharony:2008gk} and as we will see. It is straightforward to verify that the D2 brane charge is $N$ and the D6 brane charge is $N/\lambda =k$ as required.

Let us now turn to supersymmetric string configurations in this background which have finite classical action. Unlike the five-dimensional example explored in section \ref{ssec:5dsym ST}, where the string is  dynamically stabilized by the presence of a two-form flux, here instead the string must wrap a topologically non-trivial cycle. The first Betti number of $\CP^3$ is one and so there is only one possible cycle $\CP^1 \subset \CP^3$ that the string can wrap. This embedding was previously studied in \cite{Cagnazzo:2009zh} where it was shown to be supersymmetric and some of its properties were analyzed.

Remarkably, the position of the string both in the AdS$_4$ directions, as well as in the $\CP^3$ directions transverse to $\CP^1$ is completely unfixed.  As we will see momentarily and was already discussed in \cite{Cagnazzo:2009zh}, this freedom for the classical string to move gives rise to zero-modes in the spectrum of the one-loop theory. The metric on the worldsheet is found to be \eqref{sphericalworldsheet} with $L^2=\pi\ell_s^2\sqrt{2\lambda}$. The classical worldsheet action is therefore easily evaluated
\begin{equation}
\mathcal A = 2\pi \sqrt{2\lambda}\,,\qquad \text{and}\qquad \f{i}{2\pi\ell_s^2}\int B_2 = -\f{ 2\pi i l}{k} + \rmi \pi \,.
\end{equation}
When including the $B_2$-field, it is clear that there are in fact two possible string solutions for the two possible signs of the $B_2$-field coupling of the string, these are the string and anti-string solutions respectively
\be
S_\text{cl} =  2\pi \sqrt{2\lambda}\pm  \left(\f{ 2\pi i l}{k} - \rmi \pi \right)\,.
\ee
This matches precisely the field theory prediction \eqref{eq:field theory instanton ABJM}. Higher instantons can be generated by wrapping multiple strings and/or anti-strings around curves in $\CP^3$ in a similar fashion and we can convince ourselves that higher instanton terms will scale with $\e^{-2\pi n\sqrt{2\lambda}}$. The possible wrappings of strings and anti-strings in $\CP^3$ gets quite complicated to enumerate as we include higher and higher instantons and so we will start by focusing on the single instanton level in this subsection. We will comment further on higher instanton terms in the next subsection.

\begin{table}[ht]
	\begin{center}
		{\renewcommand{\arraystretch}{1.1} 
		\begin{tabular}{@{\extracolsep{10 pt}}l l c c c c}
			\toprule
			Field&Degeneracy&  $M^2L^2$ & $q$&  $\mu$&  $p$ \\
			\midrule
			\noalign{\smallskip}
			scalars& 4&$0$&0&$\frac14$ &$\frac12$\\
			&2&$-\frac12$&1&1 &$1$\\
			&2&$-\frac12$&$-1$&1 &$1$\\
			\midrule
			fermions&4&$-1$&0&1&$0$\\
			&2&$0$&1&$\frac14$&$\frac12$\\
			&2&$0$&$-1$&$\frac14$&$\frac12$\\
			\bottomrule
		\end{tabular}}
		\caption{\label{ABJMspectrum table}The spectrum of fluctuations of string modes around the classical instanton solution in the geometry \eqref{Eq: ABJM background metric}.}
	\end{center}
\end{table}
Our remaining goal is to determine the one-loop partition function \eqref{eq:1-loop pf def} for the pair of single string instantons and compare with the field theory prediction. The first factor, namely the Fradkin-Tseytlin term is easily evaluated 
\begin{equation}
	\rme^{-S_{\text{FT}}} = \frac{N^2}{\pi (2\lambda)^{5/2}}\,.
\end{equation}
Next, we need to  compute the masses of the scalar and fermion fluctuations on the worldsheet. Due to the non-trivial embedding of the string worldsheet inside $\CP^3$, the normal bundle is twisted on top of the worldsheet leading to a natural monopole field configuration \eqref{eq:monopolefield} with respect to which some of the fields are charged. 
The complete spectrum of one-loop fluctuations is independent of the $B_2$ charge of the string and is listed in table \ref{ABJMspectrum table} for both possible worldsheet instantons.\footnote{Due to the monopole we have to also include the fermionic eigenvalues linear in $m_1$ and $m_2$ (see eq. \eqref{fermionsum}), instead of solely on $M^2L^2$. These modes are, however, zero-modes and thus will have to be dealt with on their own entirely.} As anticipated, we have four massless scalars corresponding to the movement of the string in the AdS$_4$ direction leading to four obvious bosonic zero-modes. Furthermore we also find a pair of zero-modes for each of the charged scalar mode. In total we therefore have twelve bosonic zero-modes. The spectrum similarly reveals twelve fermionic zero-modes. These were already discussed in \cite{Cagnazzo:2009zh}. 
Before analyzing the zero-mode partition function, let us first evaluate the contribution of all other modes. 
To this end we can simply use the results of section \ref{ssec:oneloop} which boils down to
\begin{equation}\label{ABJMcancellation}
	-\f12 \log ({\text{Sdet}' \mathbb{K}}) = 2s_1(1) + 2 s_{1/2}(1/4) - 2 s_{1/2}(1/4) - 2s_{1}(1) = 0\,,
\end{equation}
where the fermion partition function exactly cancels the bosonic partition function. 
For the uncharged fermions in table \ref{ABJMspectrum table}, the starting value in the sum \eqref{spdef} is $p=0$ but with zero multiplicity, hence the first non-trivial contribution to the sum is from $l=1$, which is why we write $s_1(1)$ in the equation above.

Let us now turn to the contribution of the zero-modes. Usually when we encounter fermionic zero-modes, the partition function itself is not a good observable and we have to consider insertions of (fermionic) operators that soak up the zero-modes. However, in our case each fermionic zero-mode is paired with a bosonic zero-mode. A similar situation was encountered in \cite{Guica:2007wd} in the study of the extremal AdS$_2\times S^2$ `attractor' black hole in IIA string theory where it was emphasized that the combination of bosonic and fermionic zero-mode can indeed yield a finite result.
In order to work out the finite contribution of the zero-modes, the string worldsheet theory was deformed by a $Q$-exact term which localized the worldsheet to the center of AdS$_2$ and the two poles of $S^2$. As a result the zero-modes are lifted and the partition function can be evaluated.

We expect that also in our case such a localization should be possible and moreover, it should be similar in spirit to the AdS$_2\times S^2$ case. Namely, we expect the worldsheet to localize to the center of AdS$_4$ and two points on $\CP^3$. In this paper we will not perform the explicit localization computation to verify this expectation, but instead, we 
will deform the background geometry in such a way that all the zero-modes are lifted. 
The deformation will `localize' the worldsheet instanton to the center of AdS$_4$ and two possible string configurations will be allowed for the string and the anti-string, corresponding to two fixed locations in the internal space, leaving no zero-modes. 
Furthermore, the partition function can now be re-evaluated with all zero-modes lifted, and we find that there is an exact cancellation between the bosonic and fermionic fluctuations. However, due to the presence of two localization saddles, we effectively find that the zero-mode partition function is 
\be\label{zeromodePF}
Z_\text{zero-modes}=2\,.
\ee
In the section \ref{sec:3dCS} we will derive this result but for now, we will simply use it to complete the computation of the string and anti-string partition function.

Combining all ingredients we find the single instanton contribution to the partition function is (cf. \eqref{Eq:Holography} and \eqref{eq:1-loop pf def})
\begin{equation}
Z_\text{inst}^{(1)}=-\frac{\mathcal A}{8\pi^2} \frac{N^2}{\pi (2\lambda)^{5/2}} 2\rme^{-\mathcal A}\,2\cos \f{2\pi l}{k} = -  \frac{N^2}{(2\pi \lambda)^2} \rme^{-2\pi \sqrt{2\lambda}}\cos \f{2\pi l}{k}\,,
\end{equation}
in perfect agreement with the field theory answer in equation \eqref{eq:field theory instanton ABJM}. Note that crucial factors of 2 come from (1) the zero-mode partition function, (2) rewriting the contribution of string and anti-string in terms of a cosine, and (3) the measure factor $C(2)$ which we fixed already in section \ref{sec:5dsym}. The match we find here is therefore a highly non-trivial check of our value for $C(2)$ in \eqref{C2eq}.
%

\subsection{Higher instantons}
\label{ssec:higher instantons ABJM}
In this section we study the higher instanton corrections in the ABJ(M) theory. As we argued in section \ref{ssec:Higherinsta}, and confirmed in section \ref{Sec: higher instanton 5d SYM} for the IIA theory dual to 5d SYM,  higher instanton corrections coming from $n$ worldsheets overlapping on the same cycle simply get an additional factor $n^{-3}$ compared to the single instanton contribution. Combining both the strings and anti-strings we find that the higher level instantons coming from worldsheets wrapping the same cycle in our current background are given by
\begin{equation}
	Z_{\text{inst}}^{(n)} = (-1)^n\frac{N^2}{(2\pi \lambda)^2 n^3} \text{e}^{- 2\pi n \sqrt{2\lambda}} \cos \frac{2\pi n l}{k}\,,
\end{equation}
making the full sum of these instantons results in
\begin{equation}\label{Eq: simple higher instanton ABJM}
	\sum\limits_{n=1}^{\infty} Z_{\text{inst}}^{(n)} = \frac{N^2}{2(2\pi \lambda)^2} \left( \text{Li}_3(-\text{e}^{-2\pi i l/k - 2\pi \sqrt{2\lambda}}) + \text{Li}_3(-\text{e}^{2\pi i l/k - 2\pi \sqrt{2\lambda}}) \right)\,.
\end{equation}
Comparing to field theory results, however, we find that this answer does not include all contributions of higher level instantons. In \cite{Matsumoto:2013nya,Honda:2014npa,Grassi:2014uua,Kallen:2013qla} it was found that the higher instanton corrections in the ABJ(M) theory are determined by Gopakumar-Vafa (GV) invariants on $\mathbb{F}_0$. Explicitly, using the conventions of \cite{Grassi:2014uua}, it is easy to write out the contributions of higher level instantons in the type IIA limit (at large $N$ and $k$) 
\begin{equation}
	F_{\text{inst}} = \frac{N^2}{(4\pi\lambda)^2}  \sum\limits_{dm = n}   \sum\limits_{d_1+d_2 = d} \frac{(-1)^n}{n^3} n^{d_1,d_2}_0 \beta^{\frac{d_2-d_1}{d} n } \rme^{-2\pi n \sqrt{2\lambda}}\,,
\end{equation}
where the coefficients $n^{i,j}_0$ are the GV invariants on $\mathbb{F}_0$, and $\beta = \rme^{-2\pi i l/k}$.
Focusing on the tower of instantons for which $d_1 + d_2 = 1$ we find the following contributions
\begin{equation}
	  \begin{aligned}
	  	\frac{N^2}{2(2\pi\lambda)^2}   \sum\limits_{n}  \frac{(-1)^n n^{1,0}_0}{n^3} \cos \frac{2\pi n l}{k}& \rme^{-2\pi n \sqrt{2\lambda}} = \\
	  	 -\frac{N^2}{2(2\pi \lambda)^2} &\left( \text{Li}_3(-\text{e}^{-2\pi i l/k - 2\pi \sqrt{2\lambda}}) + \text{Li}_3(-\text{e}^{2\pi i l/k - 2\pi \sqrt{2\lambda}}) \right)\,,
	  \end{aligned}
\end{equation}
confirming our string theory result in \eqref{Eq: simple higher instanton ABJM}. The additional contributions from $d_1+d_2 > 1$ are presumably contributions from BPS stacks of (anti-)strings wrapping more complicated cycles in $\mathbf{C}P^3$.\footnote{Note that for the higher instantons described in section \ref{Sec: higher instanton 5d SYM} we did not find such complications because we computed the partition function in a limit for which only strings wrapping the same two-cycle contributed.} For instance, at level two we find that the instanton partition function contains a contribution from two (anti-)strings wrapping $\mathbf{C}P^1\subset \mathbf{C}P^3$, captured by the $n^{1,0}_0$-term, but also there seems to be a contribution coming from a combination of a string and an anti-string wrapping $\mathbf{C}P^1 \times \mathbf{C}P^1 \subset \mathbf{C}P^3$, coming from the $n^{1,1}_0$-term. Determining the set of all allowed supersymmetric two-cycles that the (anti-)strings can wrap requires computing the analog of Gopakumar-Vafa invariants of $\mathbf{C}P^3$, which of course is not a Calabi-Yau manifold. We are unaware of the study of such invariants in the literature, but to determine them a full supersymmetry analysis of multiple (anti-)strings in projective spaces will be necessary.


\section{Instanton corrections in 3d \texorpdfstring{$\mathcal N=2$}{N=2}  and \texorpdfstring{$\mathcal N=4$}{N=4}  CS theories}
\label{sec:3dCS}

In this section we will study worldsheet instanton corrections in two additional classes of three-dimensional supersymmetric CS theories, generalizing the results related to the ABJ(M) theory presented in the previous section. The first class arises from a one parameter mass deformation of the ABJ(M) theory \cite{Jafferis:2010un,Freedman:2013ryh}, which we will refer to as mABJ(M). The second class arises from M2 branes probing $\C^4/(\Z_r \times \Z_r)/\Z_k$. These CS theories are circular quivers consisting of $r$ copies of the ABJM quiver $\U(N)_k\times \U(N)_{-k}$ \cite{Benna:2008zy,Imamura:2008nn,Terashima:2008ba}, which we will denote as ABJM$/r$.

\subsection{Mass deformed ABJ(M) theory}
In this section we will discuss a supersymmetric deformation of ABJ(M) theory on the round $S^3$. This will lift the zero-modes encountered in section \ref{sec:ABJM} and allow us to compute the string partition function in a straightforward manner. We  start with a brief discussion of the field theory, we will closely follow the conventions of \cite{Freedman:2013ryh} where the mass deformed ABJM theory was studied from a holographic perspective. The mass deformation of ABJ(M) we are interested in can be implemented by allowing the bifundamental chiral multiplets to posses generic $\U(1)_R$ symmetry charge
\begin{equation}\label{Rdeformed}
		\begin{aligned}
			R[X^1] &= \frac12 + \delta_1 + \delta_2 + \delta_3 \,, \quad R[X^2] = \frac12 + \delta_1 - \delta_2 - \delta_3\,,\\
			R[Y_1] &= \frac{1}{2} - \delta_1 + \delta_2 - \delta_3\,,\quad R[Y_2] = \frac12 - \delta_1 - \delta_2 + \delta_3\,,
		\end{aligned}
\end{equation}
such that the 3D superpotential still has R-charge equal to $2$. This particular choice of R-charges deforms the ABJM Lagrangian with the terms\footnote{The radius of the three-sphere has been put to one in this equation and can be reinstated through dimensional analysis if needed.}
\begin{equation}
	\mathcal L_{\text{mABJM}} = \mathcal L_{\text{ABJM}} +    \sum\limits_{i=1}^3 \left[(\delta_i - 2 \prod\limits_{j\neq i}\delta_j ) \mathcal O_B^i +  \delta_i \mathcal O_F^i - \delta^2_i \mathcal O_S \right]\,,
\end{equation}
where $\mathcal O_B^i$ and $\mathcal O_S$ are scalar bilinear operators, while $\mathcal O_F^i$ are fermionic bilinears. Explicit form of the operators in terms of the chiral fields can be found in \cite{Freedman:2013ryh}.
This deformation preserves ${\cal N}=2$ supersymmetry and  $\U(1)^4$ of the original R-symmetry in the ABJM theory. To leading order in $N$ the free energy of the deformed theory is given by \cite{Jafferis:2011zi} 
\begin{equation}
	F_{\text{mABJM}} = \frac{4\sqrt{2}\pi k^{1/2}N^{3/2}}{3} \sqrt{R[X^1] R[X^2] R[Y_1] R[Y_2]}\,.
\end{equation}
Apart from the leading order answer, much less is known about the supersymmetric partition function of this theory on $S^3$, compared to the conformal ABJM theory. In \cite{Nosaka:2015iiw} it was shown that, when one of the deformation parameters vanishes, $\delta_3 = 0$, the $1/N$ perturbative contributions to the partition function still sum to an Airy function. Additionally, the author of \cite{Nosaka:2015iiw} was able to compute the first non-perturbative contributions to the partition function in a small $k$ expansion. Partly inspired by this result, it was conjectured in \cite{Bobev:2022jte,Hristov:2022lcw,Bobev:2022eus} that for generic $k$ and mass parameters, i.e. $\delta_3 \neq 0$, the $1/N$ perturbative contributions to the partition function are still of Airy-type. However, the instanton corrections to the partition function are generically not known for arbitrary deformation parameters.

As before, our interest will lie in the non-perturbative corrections to the supersymmetric $S^3$ free energy. Using string theory, we will give a prediction for the leading order worldsheet instanton correction to the partition function and discuss the possible structure of the higher instantons. We will perform our analysis in simplifying limits of the general deformation in \eqref{Rdeformed}. Two notable limits are of interest to us since their holographic duals are easier to construct in ten and eleven dimensions. The first limit is where all three deformation parameters are equal:
\begin{equation}\label{Eq: SU3 mass defo constraint}
	\delta\equiv \delta_1 = \delta_2 = \delta_3\,.
\end{equation}
%
This deformation preserves an $\SU(3)\times \U(1)$ symmetry\footnote{For $k=1,2$ the symmetry is enhanced to $\SU(3)\times \U(1)\times\U(1)$.} from the original R-symmetry in the ABJM theory. The ten and eleven-dimensional holographic duals can be constructed starting from the four-dimensional solution in \cite{Freedman:2013ryh} and uplifting them using formulae in \cite{Pilch:2015dwa}. We present these backgrounds in appendix \ref{11DsolsandSU3} but they do not turn out to be useful for lifting the zero-modes discussed in section \ref{ssec:ABJM ST}. Instead, we focus on an even simpler deformation where only one of the three $\delta$'s is non-trivial;
\be\label{SO4deltas}
\delta\equiv \delta_1\,,\qquad \delta_2 = \delta_3 =0\,.
\ee
In this case, the symmetry is enhanced to $\SO(4)\times\U(1)$.\footnote{For $k=1,2$ the symmetry is enhanced to $\SO(4)\times \SO(4)$.} In the next subsection we will present the ten-dimensional geometry dual to this deformation and consider worldsheet instantons in that background. We will find that the zero-modes of the string can be lifted in such a way that a finite value \eqref{zeromodePF} can be assigned to the zero-mode partition function by taking the $\delta\to0$ limit. Furthermore, we can give a prediction for the worldsheet instanton contribution to the partition function for generic values of $\delta$. 

A subtle point we would like to emphasize before describing the ten-dimensional uplifts of the mass deformations is that all deformations coming from \eqref{Rdeformed} come in pairs of two branches. The difference between the two branches is given by the choice of Killing-spinor \cite{Freedman:2013ryh}, and fixes how the $S^3$ isometry group $\SO(4) = \SU(2)_\ell \times \SU(2)_r$ is embedded in the full preserved supergroup. In particular, one branch preserves $\SU(2)_\ell \times \OSp(2|2)_r$, while the other preserves $\OSp(2|2)_\ell \times \SU(2)_r$. From now one we will label these branches mABJ(M)$^+$ and mABJ(M)$^-$ respectively.

\subsection{Single mass deformation of ABJ(M)}\label{ssec:singlemassdeform}
In this section we will uplift the $\SO(4)\times\SO(4)$ invariant solution of \cite{Freedman:2013ryh}. In their language this is the solution for which $c_2 = c_3 =0$, but $c_1\equiv c \ne0$. This choice of parameters exactly corresponds to the R-charge deformation \eqref{SO4deltas} and in this case $c=\delta/2$ \cite{Freedman:2013ryh}. The eleven-dimensional solution is a straight-forward application of the uplift formulae in \cite{Cvetic:1999au} and can be found in appendix \ref{11DsolsandSU3} (see also \cite{Bobev:2013yra} where the uplift was performed for a similar application). Once the eleven-dimensional solution is found, we can perform a direct dimensional reduction to ten dimensions. It turns out that the only difference between the two branches of mass deformations mABJ(M)$^\pm$ is encoded in the form fields, as we will see shortly. The metric is a direct product of the round metric on AdS$_4$ and a squashed metric on $\CP^3$ where the squashing depends on the radial variable in AdS$_4$. We will write the explicit metric on AdS$_4$ as before in \eqref{AdS4metric}. Using this we find the squashed metric on $\CP^3$ is
\be\label{CP3so4}
\dd s_{6}^2 = \dd\theta^2 + \f{\cos^2\theta}{4 Y_1}  (\dd \theta_1^2+\sin^2\theta_1\dd\phi_1^2)+ \f{\sin^2\theta}{4 Y_2}  (\dd \theta_2^2+\sin^2\theta_2\dd\phi_2^2) + \sin^2\theta\cos^2\theta\,\Sigma^2
\ee
where 
\be
\Sigma = \dd \varphi + \f12 \cos\theta_1\,\dd\phi_1 - \f12 \cos \theta_2\,\dd\phi_2\,.
\ee
The two functions $Y_1$ and $Y_2$ implement the squashing of the internal space and take the form
\be
Y_1=1+c \f{\cos^2\theta}{\cosh^2(\rho/2)}\,,\qquad Y_2=1-c \f{\sin^2\theta}{\cosh^2(\rho/2)}\,,
\ee
where the parameter $c$ controls the squashing. Notice that the squashing functions depend on the radial variable of AdS$_4$. This means that, even though, naively the four-dimensional metric exhibits an $\SO(5,1)$ isometry, the full ten-dimensional metric does not. The ten-dimensional metric preserves an $\SO(4)$ isometry generated by the rotations of the three-sphere in AdS$_4$, an $\SU(2)\times\SU(2)$ generated by rotations of the two 2-spheres, and a $\U(1)$ generated by a shift of the angle $\varphi$. The metric \eqref{CP3so4} on $\CP^3$ is slightly non-standard, and so we recall some quantities of the undeformed metric ($c\to0$) in our coordinates. The K{\"a}hler form of the undeformed metric can be found as $J=\dd\omega$ where
\be
\omega = \f14 (\cos(2\theta)\dd\varphi+\cos^2\theta\cos\theta_1\dd\phi_1+\sin^2\theta\cos\theta_2\dd\phi_2)\,.
\ee
It is easy to verify that the volume form on the undeformed space is $J^3/3!$, using the fact that $\varphi$ is $2\pi$-periodic and $0<\theta<\pi/2$ we obtain that
\be
\int \f{J^3}{3!} = \f{\pi^3}{3!}\,,
\ee
as required. The ten-dimensional metric and dilaton can now be written as
\be
\begin{split}
\dd s_{10}^2&=L^2(\dd s_{\text{AdS}_4}^2 + 4 \dd s_{6}^2)\,,\\
\e^{2\Phi} &= \f{\pi(2\lambda)^{5/2}}{N^2 Y_1 Y_2}\,,
\end{split}
\ee
while the form fields are
\be
\begin{split}
C_1 &= \f{N \ell_s}{\lambda}\Big(2\omega -\f{c\sin^2\theta\cos^2\theta}{\cosh^2(\rho/2)}\Sigma\Big)\,,\\
B_2 &= \pm\f{i \pi c \ell_s^2\sqrt{2\lambda}}{\cosh^2(\rho/2)}\Big( -\f{\cos^4\theta}{Y_1}\vol_{S^2_1} +\f{\sin^4\theta}{Y_2}\vol_{S^2_2} \Big) + \left(\f12 -\frac{ l }{k}\right) 4\pi\ell_s^2 J\,,\\
C_3 &=\f{N i\pi \ell_s^3}{\sqrt{2\lambda}} \bigg(-3U \vol_{S^3} \pm \f{c }{\cosh^2(\rho/2)}\dd\varphi\w \Big( \f{\cos^4\theta}{Y_1}\vol_{S^2_1} +\f{\sin^4\theta}{Y_2}\vol_{S^2_2} \Big)\bigg)\,,
\end{split}
\ee
where the choice of signs depends on the mABJ(M)$^{\pm}$ branch of mass deformation chosen. In this expression we have introduced the volume forms of the unit-radius spheres. The volume form on the $S^3$ slices of AdS$_4$ is denoted by $\vol_{S^3}$, the volume form on $S^2_1$ spanned by coordinates $(\theta_1,\phi_1)$ is denoted by $\vol_{S^2_1}$, and the volume form on $S^2_2$ spanned by the coordinates $(\theta_2,\phi_2)$ is denoted by $\vol_{S^2_2}$. 
Furthermore, the function $U$ is given by
\be
U = \f{1}{12}\Big( -9\cosh\rho + \cosh(3\rho) + 16 c \cos(2\theta)\sinh^4(\rho/2) \Big)\,,
\ee
and it reduces to \eqref{Uundeformed} in the $c\to0$ limit. 
Notice that we have added a term to the $B_2$ form field along $J$ to incorporate the ABJ theory as before. All quantized fluxes are the same as for the undeformed metric which means that our ten-dimensional solution describes the holographic dual to $\U(N+l)_k\times\U(N)_{-k}$ ABJ theory deformed by a mass parameter related to $c$.

We can use the standard formula to compute the leading order contribution to the on-shell supergravity action, by relating the effective four-dimensional Newton constant $G_{(4)}$ to the ten-dimensional Newton constant times the volume (weighted by the dilaton) of the six-dimensional space 
\be
S_\text{sugra} =\f{\pi L^2}{2 G_{(4)}} = \f{16\pi^3L^2}{(2\pi \ell_s)^8} \int\e^{-2\Phi}\sqrt {g_6}  = \f{2\pi N^2}{3\sqrt{2\lambda}}\,.
\ee
This expression is independent of the deformation parameter $c$ in agreement with \cite{Freedman:2013ryh}. However, the free energy of mass deformed ABJ(M) does depend on the parameter $c$ in a non-trivial manner, even at leading order. The discrepancy comes from the fact that we still have to Legendre transform the on-shell action to account for the fact that the deformation parameter $c$ is really a source for a scalar that is subject to alternative quantization. This procedure was performed in \cite{Freedman:2013ryh} resulting in the leading order free energy
\be
{\cal F} = \f{2\pi N^2}{3\sqrt{2\lambda}}(1-c^2)\,,
\ee
matching the field theory prediction \cite{Jafferis:2011zi} after identifying $\delta$ with $2c$.

We will now analyze worldsheet instantons in the two backgrounds described above. In all cases, the string (or anti-string) wraps a diagonal combination of the two 2-spheres $S^2_1$ and $S_2^2$. It is easy to compute the classical area and $B$-field contribution to the string (and anti-string) action to be\footnote{This classical contribution to the partition function is consistent with the prediction on the QFT side in \cite{Nosaka:2015iiw}, when the identification $\xi=\eta\equiv c$ is made in that paper. We are thankful to 
Tomoki Nosaka for pointing this out.}
\begin{equation}
\mathcal A = 2\pi \sqrt{2\lambda}(Y_1Y_2)^{-1}\,,\qquad \text{and}\qquad \mathcal B=\f{i}{2\pi\ell_s^2}\int B_2 = \pm 2\pi\sqrt{2\lambda}\big(1-(Y_1Y_2)^{-1}\big)-\f{ 2\pi i l}{k} + \rmi \pi\,,
\end{equation}
where again the sign is determined by the choice of mass deformation. For the moment focussing on the mABJ(M)$^{+}$ branch we see that for the string, that is when we add the terms ${\cal A}$ and ${\cal B}$, the two $c$-dependent contributions cancel and we recover the classical action we obtained for the undeformed geometry. In this case the classical position of the string is also not fixed since all $\rho$-dependence and dependence on the remaining $\CP^3$ coordinates drops out. When analysing the spectrum we recover exactly the same spectrum as for the undeformed theory, cf. table \ref{ABJMspectrum table}, which is  again plagued by zero-modes. On the other hand for the anti-string, the two $c$-dependent terms add up and we must extremize the resulting function. We find two extrema, the first one at 
\be\label{ext1}
\rho=0 \quad \text{and}  \quad \theta=\pi/2\,,
\ee 
and the second one at 
\be\label{ext2}
\rho=0  \quad \text{and}  \quad \theta=0\,.
\ee
A similar analysis can also be done for the mABJ(M)$^-$ branch. In this case the situation is exactly reversed, namely here we find that the string is stabilized to two fixed positions (\eqref{ext1} and \eqref{ext2}), whereas the anti-string is left unfixed with its associated zero-modes. We summarize the classical on-shell action of the (anti-)strings, and their number of zero-modes, for both mABJM$^{\pm}$ branches in table \ref{massdeformedstringsols}.

\begin{table}[ht]
	\begin{center}
		{\renewcommand{\arraystretch}{1.1} 
			\begin{tabular}{@{\extracolsep{10 pt}}l c c c}
				\toprule
				Branch&string charge&  $S_\text{cl}$ & \# zero-modes \\
				\midrule
				\noalign{\smallskip}
				& $\pm1$&$2\pi\sqrt{2\lambda} \mp 2\pi i ( \frac{l}{k} - \frac12)$&12\\
				mABJ(M)$^{\pm}$& $\mp1$&$2\pi \sqrt{2\lambda}\f{1+c}{1-c} \pm 2\pi i ( \frac{l}{k} - \frac12)$&0\\
				& $\mp1$&$2\pi \sqrt{2\lambda}\f{1-c}{1+c} \pm 2\pi i ( \frac{l}{k} - \frac12)$&0\\
				\bottomrule
		\end{tabular}}
		\caption{\label{massdeformedstringsols}The possible string and anti-string solutions in the mass-deformed ABJ(M) geometry together with their classical action. For each branch mABJ(M)$^{\pm}$ we find a pair of stabilized (anti-)strings and an unfixed (anti-)string. The unstabilized string behaves exactly as in the undeformed ABJ(M) geometry discussed in section \ref{ssec:ABJM ST}, namely the spectrum and number of zero-modes is unchanged.}
	\end{center}
\end{table}

By using the two branches in tandem we can now compute both the string and anti-string partition functions: 
We are going to use the mABJ(M)$^-$ branch in order to lift the string zero-modes present in the undeformed ABJ(M) theory, while we are going to utilize the mABJ(M)$^+$ branch to eliminate the anti-string zero modes of the undeformed case. 
Since there are no zero-modes for the anti-string in the mABJ(M)$^+$ branch and the string in the mABJ(M)$^-$ branch, we are able to compute the partition functions in a straightforward manner. 
In order to obtain the partition functions of strings and anti-strings with zero-modes, we simply tune our deformation parameter $c$ back to zero, where all the zero-modes re-emerge. By assuming that this limit is smooth, we can safely assign a value to the zero-mode partition function for the string and anti-string. 
As a by-product, this also allows us to give a prediction for the instanton contributions to the partition function of the mass-deformed ABJ(M) theory.

Let us then go ahead and compute the partition function for the string and anti-string in the mABJ(M)$^-$ and mABJ(M)$^+$ branches respectively. We start by noticing that the combination of the measure factor $C(2)$ and the dilaton contribution is the same for all cases, and it is independent of $c$:
\be
C(2)\e^{-S_\text{FT}} = \f{\cal A}{8\pi^2}\e^{-2\Phi} =  \f{N^2 }{16\pi^2\lambda^{2}}\,.
\ee
The entire $c$-dependence of the string partition function therefore comes from the fluctuations of the string modes. 

Our next task is to compute the spectrum of the deformed (anti-)strings. As explained above, the classical action of the string or anti-string is not affected by the deformation parameter $c$, depending on the sign choice in mABJ(M)$^{\pm}$, hence the corresponding fluctuations for these strings have exactly the same spectrum as in the undeformed geometry, i.e. we recover exactly table \ref{ABJMspectrum table}. For the other string, however, the spectra are modified. In table \ref{deformedABJMspectrum table}, we report the spectrum corresponding to string fluctuations around the classical solution \eqref{ext2}, where the upper sign in all entries is associated to the anti-strings in the mABJ(M)$^{+}$ deformation, while the lower signs are associated to strings in the mABJ(M)$^{-}$ deformation.

\begin{table}[ht]
	\begin{center}
		{\renewcommand{\arraystretch}{1.1} 
		\begin{tabular}{@{\extracolsep{10 pt}}l l c c c c}
			\toprule
			Field&Degeneracy&  $M^2L^2$ & $q$&  $\mu$&  $p$ \\
			\midrule
			\noalign{\smallskip}
			scalars& 4&$\f{\pm c}{(1\pm c)^2}$&0&$\frac14 - \f{\pm c}{(1 \pm c)^2}$ &$\frac12$\\
			&2&$\frac12-\frac{1}{(1 \pm c)^2}$&1&$\frac{1}{(1 \pm c)^2}$ &$1$\\
			&2&$\frac12-\frac{1}{(1\pm c)^2}$&$-1$&$\frac{1}{(1\pm c)^2}$ &$1$\\
			\midrule
			fermions&4&$-\frac{1}{(1 \pm c)^2}$&0&$\frac{1}{(1 \pm c)^2}$&$0$\\
			&2&$\f{\pm c}{(1\pm c)^2}$&1&$\frac14-\f{\pm c}{(1 \pm c)^2}$&$\frac12$\\
			&2&$\f{\pm c}{(1 \pm c)^2}$&$-1$&$\frac14-\f{\pm c}{(1 \pm c)^2}$&$\frac12$\\
			\bottomrule
		\end{tabular}}
		\caption{\label{deformedABJMspectrum table}The spectrum of fluctuations of string modes around the classical instanton solution in the single mass deformed ABJ geometry. The spectra shown correspond to the (anti-)strings with classical action equals to $2\pi \sqrt{2\lambda}\f{1\mp c}{1 \pm c}$.}
	\end{center}
\end{table}
The remarkable feature of the spectrum in table \ref{deformedABJMspectrum table} is that the degeneracies are exactly as for the undeformed ABJ(M) instantons. Moreover, the deformation respects the splitting among the different modes as it was for the undeformed case, in such a way that the shifted masses $\mu$ for uncharged (charged) fermions and charged (uncharged) bosons are equal. 
Since we do not have any zero-modes for generic non-zero values of $c$,%
\footnote{We are interested in small non-zero values of $c$ around the value $c=0$. Each of two extrema \eqref{ext1} and \eqref{ext2} impose symmetric (in $c\to-c$) restrictions on the interval of the mass deformation parameter in order to avoid zero-modes. As a result, we can safely consider $-\f 12 <c <\f 12$ and $c\neq 0$.}
this means that, except for the lowest level fermionic states that give contributions that depend linearly on $m_1$ and $m_2$, there is an exact cancellation between fermions and bosons as we found for the non zero-modes in the undeformed theory \eqref{ABJMcancellation}. The eigenvalues of the lowest weight states are given by
\begin{equation}
m_1 L = \f12\text{sgn}(q)\,,\qquad m_2 L = \f{1\mp c}{2(1\pm c)}
\end{equation}
where the $\pm$ signs are correlated exactly  as in table \ref{deformedABJMspectrum table}. Here $\text{sgn}(q)$ refers to the sign of the charge of the relevant fermion which can also be read from table \ref{deformedABJMspectrum table}. The contributions of the quantum fluctuations for each one of the deformed and stabilized (anti-)strings is therefore
\be\label{mABJcancellation}
\begin{split}
	-\f12 \log ({\text{Sdet} \mathbb{K}}) 
	&= 2s_1\left(\frac{1}{(1 \pm c)^2}\right) + 2 s_{1/2}\left(1/4-\f{\pm c}{(1 \pm c)^2}\right) 
	\\ 
	&- 2 s_{1/2}\left(1/4-\f{\pm c}{(1 \pm c)^2}\right)  - 2s_{1}\left(\frac{1}{(1 \pm c)^2}\right) +  \log (1\pm c)^2 = \log (1\pm c)^2\,,
	\end{split}
\ee
where the final contribution purely arises from the lowest weight states, as given in equation \eqref{fermionsum} and Appendix \ref{App: Fermionic monopole harmonics}. The conclusion is thus that the anti-string partition function of the mABJ(M)$^+$ branch and the string partition function of the mABJ(M)$^-$ branch consists each of two saddles. 
These two saddles contribute in a similar manner to the partition function, differing by the sign of $c$, and when $c$ is taken to vanish their contribution is simply $1$.
Hence, we can safely conclude that the zero-mode partition function we encountered in the undeformed theory simply counts how many saddles the string (or anti-string) has in the deformed theory. 
Our computation shows that for either the string or the anti-string there are two saddles when they are deformed and so we conclude that 
\be\label{zeromodePF2}
Z_\text{zero-modes}=2\,.
\ee

Armed with this result, we can now also give a prediction for the first instanton correction of the single mass deformed ABJ(M) theory for the two branches:
\begin{equation}\label{zinstaso41}
	\text{mABJ(M)}^\pm\,:\quad Z_\text{inst}^{(1)} =  - \frac{N^2}{(4\pi \lambda)^2} \bigg(2\text{e}^{-2\pi \sqrt{2\lambda} \pm \frac{2\pi i l}{k}}  +(1-c)^2\text{e}^{-2\pi \sqrt{2\lambda}\frac{1+c}{1-c} \mp \frac{ 2\pi i l}{k}}+(1+c)^2\text{e}^{-2\pi \sqrt{2\lambda}\frac{1-c}{1+c} \mp \frac{ 2\pi i l}{k}} \bigg)\,.
\end{equation}
Applying our results for the wrapping of multiple strings in section \ref{ssec:Higherinsta}, we predict the existence of a tower of higher level worldsheet instantons in the mass deformed theories that can be resummed into
\begin{equation}
	 \begin{aligned}
	 	\frac{N^2}{(4\pi\lambda)^2} \Bigg( 2 \,\text{Li}_3 \left[ -\rme^{-2\pi \sqrt{2\lambda} \pm \frac{2\pi\rmi l}{k}} \right]& +(1-c)^2 \text{Li}_3 \left[ -\rme^{-2\pi \sqrt{2\lambda} \frac{1+c}{1-c} \mp \frac{2\pi\rmi l}{k}} \right]  \\
	 	&+(1+c)^2 \text{Li}_3 \left[ -\rme^{-2\pi \sqrt{2\lambda} \frac{1-c}{1+c} \mp \frac{2\pi\rmi l}{k}} \right] \Bigg)\,.
	 \end{aligned}
\end{equation}
At this stage we note that in order to give a definite prediction for the first instanton correction to the free energy of the mass deformed ABJ(M) we may have to perform the Legendre transform to account for alternative quantization (see \cite{Freedman:2013ryh,Bobev:2018wbt} for the relevant computation in our context). It is not clear how to proceed with it using our approach, since we should vary the instanton partition function with respect to the na\"ive source as an intermediate step. Performing this variation is however not practical since it implies varying the quantum effective action of the string. A more appropriate approach would be to compute the instanton correction to 4D supergravity and phrase the computation entirely in that language. 
 It would be interesting to either explicitly work out the instanton corrected 4D supergravity theory in order to facilitate the comparison to field theory or to perform a direct field theory computation of the instanton effects for finite mass parameter.

\subsection{Orbifolds of the ABJM theory}
\label{sec:orbifolds-abjm}

As mentioned at the beginning of this section, the second class of examples we will study are $\mathcal N=4$ circular quivers constructed from $r$ copies of the ABJM theory with finite CS level $k$.  In the large $k$ limit, these are realized in type IIA string theory as the theory of D2 branes probing the orbifold $(\C^4/\Z_r\times \Z_r)/\Z_k$, where the orbifold action is realized as followed \cite{Imamura:2008nn}
\begin{equation}
	\begin{aligned}
		&(z_1,z_2 ,z_3,z_4) \rightarrow (\rme^{\frac{2\pi\rmi}{r}}z_1,\rme^{-\frac{2\pi\rmi}{r}}z_2 ,z_3,z_4)\,,\\
		&(z_1,z_2 ,z_3,z_4) \rightarrow (z_1,z_2 ,\rme^{\frac{2\pi\rmi}{r}}z_3,\rme^{-\frac{2\pi\rmi}{r}}z_4)\,,\\
		&(z_1,z_2 ,z_3,z_4) \rightarrow (\rme^{\frac{2\pi\rmi}{kr}}z_1,\rme^{-\frac{2\pi\rmi}{kr}}z_2 ,\rme^{\frac{2\pi\rmi}{kr}}z_3,\rme^{-\frac{2\pi\rmi}{kr}}z_4)\,.\\
	\end{aligned}
\end{equation}
The perturbative contribution to the supersymmetric partition function of this theory was found to be an Airy function \cite{Honda:2014ica} in a similar fashion as the ABJM theory. At large $N$, the free energy of the orbifolded theory is the same as the free energy of the ABJM theory rescaled by the orbifold number $r$ \cite{Imamura:2008nn}:
\begin{equation}
	F_{\text{ABJM}/r}^{\text{pert}} = r F_{\text{ABJM}}^{\text{pert}}\,.
\end{equation}
This can be demonstrated explicitly from the holographic dual. In type IIA string theory the dual to the orbifolded theory is expressed as
\be
\begin{split}
\dd s_{10}^2&=L^2(\dd s_{\text{AdS}_4}^2 + 4 \dd s_{\CP^3/\Z_r}^2)\,,\\
\e^{2\Phi} &= g_s^2=\f{\pi(2\lambda)^{5/2}}{N^2r^2}\,,
\end{split}
\ee
where the length scale $L$ is related to the 't~Hooft coupling as usual $L^2= \pi \ell_s^2\sqrt{2\lambda}$.
Using this, the holographic free energy is easily computed to be
\be
S_\text{sugra} =\f{\pi L^2}{2 G_{(4)}} = \f{16\pi^3L^2}{(2\pi \ell_s)^8} \f{1}{g_s^2}\vol_6 = \f{2\pi N^2 r}{3\sqrt{2\lambda}}\,,
\ee
which indeed is the rescaled ABJM free energy by a factor $r$.

Due to the orbifold, the worldsheet instanton now wraps $\CP^1/\Z_r$ in the string theory background. The contribution of the level one worldsheet instantons to the partition function was argued to be \cite{Honda:2014ica, Hatsuda:2015lpa} 
\begin{equation}\label{orbifoldinstanton}
\frac{N^2 r^4}{(2\pi \lambda)^2} \rme^{-\frac{2\pi}{r} \sqrt{2\lambda}}\,.
\end{equation}
This result can be understood from string theory using our results. Comparing with the original ABJM computation, the orbifolding now acts on the worldsheet. Therefore the classical area is scaled down by a factor $r$. This can be observed from the exponential behaviour of the instanton contribution. This is reminiscent of our earlier discussion of higher instantons where the multiple winding of the string rescales the effective area. Notice however, that this rescaling of the area is slightly different from what we have seen in previous examples for higher instantons. For higher instantons the effective area of the worldsheet is multiplied by an integer whereas now, the $\CP^1$ is acted upon by an orbifolding of the background geometry and so its area is correspondingly scaled down ${\cal A} = 2\pi \sqrt{\lambda}/r$. Just as for higher instantons, the orbifolding of the worldsheet renders it mildly singular, namely the poles of the spheres are replaced by orbifolds with associated deficit angles. Due to the two orbifold points on the worldsheet, we must include the factors $z_{1/r}^2$ just as we did for higher instantons.  Additionally, in this case we must  also use the rescaled area in the prefactor $C(2)$, this is different to the case of higher instantons where the string fluctuations are always measured with respect to the area of the single string worldsheet. Besides these modifications, the rest of the computation proceeds effectively identically to what we saw for the ABJM geometry. In particular the (dimensionless) spectrum of string fluctuations is identical to the original ABJM spectrum. Combining the factors we find
\be
\begin{split}
Z_\text{inst}^{(1)} &= 2\,\f{\cal A}{8\pi^2}\, z_{1/r}^2\, g_s^{-2}\,(\text{Sdet}'{\mathbb{K}})^{-1/2}Z_\text{zero-modes}\e^{-{\cal A} + \pi i}\\
& =  - \frac{N^2r^4}{(2\pi\lambda)^{2}} \rme^{-\frac{2\pi}{r} \sqrt{2\lambda}}\,,
\end{split}
\ee
where the first factor of $2$ is due to the string and anti-string both contributing to the answer. In order to evaluate the second line, we used our result for the zero-mode partition function \eqref{zeromodePF2} and the non-zero mode contribution \eqref{ABJMcancellation} to find a precise match with the field theory result \eqref{orbifoldinstanton}. Following similar steps as in \ref{ssec:higher instantons ABJM} we find that there is a series of higher instanton contributions which can be summed into
\begin{equation}
	\frac{N^2 r^4}{(2\pi\lambda)^2} \text{Li}_3\left( -\rme^{-\frac{2\pi}{r}\sqrt{2\lambda}} \right)\,.
\end{equation}
These instantons arise from worldsheets wrapping the same sphere multiple times. We know, however, that there are more complicated cycles that the worldsheets can wrap, which are not incorporated in this answer. Working out the contributions of such higher instantons to the string partition function is left for future work.

\section{Summary and future prospects}
\label{sec:discussion}
In this paper we computed worldsheet instanton contributions to the string partition function in the low energy limit in a number of holographic type IIA backgrounds. Since the string partition function is sensitive to unknown measure factors, we employ holography and exact results on the QFT side obtained by  supersymmetric localization, to fix all contributions to the level one genus zero worldsheet instanton at one-loop. To summarize, we find that for genus zero
\begin{equation}\label{Eq: result for genus 0 level 1}
	Z_{\text{inst}}^{(1)} = \frac{\mathcal A}{8\pi^2} \text{e}^{-S_{\text{FT}}} (\text{Sdet}'\mathbb{K})^{-1/2} Z_{\text{zero-modes}}\,\text{e}^{-S_\text{cl}}\,,
\end{equation}
where $\mathcal A$ is the area of the worldsheet, $S_{\text{FT}}$ is the Fradkin-Tseytlin term, and the superdeterminant takes all kinetic operators of the CFT living on the worldsheet into account, excluding the zero-modes, which have to be analyzed separately and are collected in $Z_{\text{zero-modes}}$. Finally, $S_{\text{cl}}$ captures the classical contribution of the string on-shell action. Two highly non-trivial checks for the validity of this formula are the explicit matches with known QFT results in maximal SYM on $S^5$ with gauge group $\text{U}(N)$, and the ABJ(M) theory on $S^3$ with gauge group $\text{U}(N+l)_k \times \text{U}(N)_{-k}$. Specifically we find at large $N$ that
\begin{equation}
	\begin{aligned}
		&\text{5d SYM}\,:\quad Z_{\text{inst}}^{(1)} = - \frac{N^2}{\xi^2}\text{e}^{-\xi} \,,\\
		&\text{ABJ(M)}\,:\quad Z_{\text{inst}}^{(1)} = - \frac{N^2}{(2\pi\lambda)^2} \text{e}^{-2\pi \sqrt{2\lambda}} \cos \frac{2\pi l}{k}\,,
	\end{aligned}
\end{equation}
matching field theory results \cite{Kim:2012ava,Drukker:2010nc} in the type IIA limit. Furthermore, for a multi-wound string on a sphere we conjecture that its one-loop contribution is given by that of a single worldsheet instanton times a correction factor $z_n^2$ solely depending on an angular excess of $2\pi(n-1)$ on the poles, due to the multiple windings. To determine the correction we studied multi-wound strings wrapping a disk in AdS$_5 \times S^5$, dual to 1/2-BPS Wilson loops in four-dimensional $\mathcal N=4$ SYM. Comparing to localization results in the QFT \cite{Pestun:2007rz} we find that $z_n = n^{-3/2}$. Since a multi-wound string wrapping on a sphere contains two poles with excess angles, compared to a single pole when wrapping a disk, we conclude that the corresponding $n$-th level worldsheet instanton is given by
\begin{equation}\label{Eq: result for genus 0 level n}
	Z_{\text{inst}}^{(n)} = \frac{\mathcal A}{8\pi^2 n^3} \text{e}^{-S_{\text{FT}}} (\text{Sdet}'\mathbb{K})^{-1/2} Z_{\text{zero-modes}} \text{e}^{-n S_{\text{cl}}}\,.
\end{equation}
Indeed, when applying this formula to the five-dimensional SYM theory we find that 
\begin{equation}\label{eq: summary for 5d SYM all instantons}
	\text{5d SYM}\,:\quad \sum\limits_{n=1}^\infty Z_{\text{inst}}^{(n)} = - \frac{N^2}{\xi^2} \text{Li}_3(\text{e}^{-\xi})\,,
\end{equation}
matching the field theory result in \eqref{5DLogZ} on the nose. For the ABJ(M) theory the result of summing these multi-wound strings is
\begin{equation}
	\text{ABJ(M)}\,:\quad \sum\limits_{n=1}^\infty Z_{\text{inst}}^{(n)} = \frac{N^2}{2(2\pi \lambda)^2} \left( \text{Li}_3(-\text{e}^{-2\pi i l/k - 2\pi \sqrt{2\lambda}}) + \text{Li}_3(-\text{e}^{2\pi i l/k - 2\pi \sqrt{2\lambda}}) \right) \,.
\end{equation}
Comparing to the field theory \cite{Matsumoto:2013nya,Drukker:2010nc,Hatsuda:2012dt} we find that this answer only partially matches. The reason is that the worldsheet instantons can wrap more complicated curves in the $\mathbf{C}P^3$ background, where our formula only covers the strings wrapping $\mathbf{C}P^1 \subset \mathbf{C}P^3$ multiple times. Determining all possible embeddings of the worldsheet instantons requires a careful analysis of the BPS conditions for multiple strings. This is an interesting mathematical problem as it requires the knowledge of Gopakumar-Vafa-like invariants for the non-Calabi-Yau manifold $\mathbf{C}P^3$.

As an application of our results we predict the worldsheet instanton corrections in two deformations of the ABJ(M) theory, for which the contributions to the dual QFT free energy is not known in the literature. 
The first example we discuss is the mass-deformed ABJ(M) theory preserving an $\text{SO}(4) \times \text{U}(1)$ internal symmetry. As discussed in \cite{Freedman:2013ryh}, there are two such branches of massive deformations, with mass parameter $c$, which we called mABJM$^\pm$. For both these branches we have computed a tower of instanton contributions that sum to
\begin{equation}\label{high-mabjm}
	\begin{aligned}
		\text{mABJM}^\pm\,:\quad \sum\limits_{n=1}^\infty & Z_{\text{inst}}^{(n)} = \, \frac{N^2}{(4\pi\lambda)^2} \bigg( 2 \,\text{Li}_3 \left[ -\rme^{-2\pi \sqrt{2\lambda} \pm \frac{2\pi\rmi l}{k}} \right]\\
	& + (1-c)^2 \text{Li}_3 \left[ -\rme^{-2\pi \sqrt{2\lambda} \frac{1+c}{1-c} \mp \frac{2\pi\rmi l}{k}} \right]  + (1+c)^2  \text{Li}_3 \left[ -\rme^{-2\pi \sqrt{2\lambda} \frac{1-c}{1+c} \mp \frac{2\pi\rmi l}{k}} \right] \bigg)\,,
	\end{aligned}
\end{equation}
where the $\pm$ signs are determined by the choice for the branch of the deformation. This may not be compared directly with the free energy of the dual QFT as we may have to Legendre transform our answer to account for alternative quantization of the scalar that is turned on in the four-dimensional supergravity solution and is controlled by the parameter $c$. Since we do not have the instanton correction to the four-dimensional supergravity (in particular the K{\"a}hler potential) it is not clear how to carry out this Legendre transform. 

The second example is an orbifold of the ABJM theory, preserving $\mathcal N=4$ supersymmetry, constructed from $r$ copies of the ABJM theory. The type IIA description arises from a stack of D2 branes probing the orbifold $(\mathbf{C}^4/\mathbf{Z}_r^2)/\mathbf{Z}_k$. The worldsheet instantons coming from strings wrapping $\mathbf{C}P^1/\mathbf{Z}_r $ multiple times in the background sum to give
\begin{equation}\label{high-orb}
	\text{ABJM}/r\,:\quad \sum\limits_{n=1}^\infty Z_{\text{inst}}^{(n)} =\frac{N^2 r^4}{(2\pi\lambda)^2} \text{Li}_3\left( -\text{e}^{-\frac{2\pi}{r}\sqrt{2\lambda}} \right)\,.
\end{equation}
A non-trivial consistency check for this result is that the first level instanton matches the QFT computation in \cite{Hatsuda:2015lpa}. It is important to remember that the expressions in \eqref{high-mabjm} and \eqref{high-orb} are not the complete answers for $n>1$, cf. discussions in sections \ref{ssec:higher instantons ABJM} and \ref{sec:orbifolds-abjm}.

As a byproduct of our holographic study of mass-deformed ABJ(M) theory we present in appendix \ref{App: Holographic duals to mass deformed ABJM} novel eleven-dimensional and type IIA supergravity backgrounds, which are dual to mass deformations of the ABJ(M) theory preserving $\text{SU}(3)\times \text{U}(1)^2$ and $\text{SU}(2)^2 \times \text{U}(1)$ internal symmetries respectively.

All of our results are directly obtained in type IIA string theory, it will be interesting to uplift this analysis to eleven dimensions. Recently, a similar study as we have presented in this paper was applied to compute the one-loop partition function of an M2-brane dual to a Wilson loop \cite{Giombi:2023vzu}. Computing the one-loop partition function of M2-brane instantons in a similar fashion will allow us to generalize our results away from the large 't Hooft coupling limit. Specifically, the instanton correction to 5D SYM computed in this paper is particularly suited for this approach when uplifted to M2-branes eleven dimensions. 
These branes wrap an $S^2 \times S^1_\beta$ in the AdS$_7 \times S^4$ M-theory background, and so is a product space just as it happened in \cite{Giombi:2023vzu}. Here $S^1_\beta \subset \text{AdS}_7$ and $S^2 \subset S^4$. The contributions of such M2-branes to the supersymmetric partition function of the $(2,0)$ theory living on the $S^5\times S^1_\beta$ boundary of AdS$_7$ was computed before using the so called giant graviton expansion \cite{Arai:2020uwd}. The partition function of these branes was computed by localizing the ABJM theory living on the M2-branes in the curved background. It will be interesting to make a direct comparison between the giant graviton expansion and our exact result in \eqref{eq: summary for 5d SYM all instantons}. Utilizing such a direct comparison, we may derive $C(\chi = 2n)$ in a similar spirit to how $C(\chi = 2n+1)$ was derived in  \cite{Giombi:2023vzu}.

The worldsheet instantons in the ABJ(M) theory on the other hand uplift to M2-branes wrapping Lens-spaces in the eleven-dimensional AdS$_4 \times S^7/\mathbf{Z}_k$ background. It was argued in the field theory that the contributions of these instantons can be summed into an Airy function \cite{Hatsuda:2012dt}. Inspired by this result, and the giant graviton expansion, one can wonder if there exists a giant instanton expansion of M2-branes encompassing the non-perturbative corrections to the ABJ(M) $S^3$ supersymmetric partition function, which can be computed by localizing the ABJM theory living on the M2-branes wrapping Lens-spaces in the curved internal geometry. Such a computation would generalize our result to any value of $k$ in the field theory, and additionally the analysis would also include the D2-brane instantons in the ABJ(M) matrix model, which we have neglected in our analysis since they are subleading in $N$ in the IIA limit. 

A final prospective application of our results lays in the context of black holes and the quantum corrections to its Bekenstein-Hawking entropy. Studying these corrections directly in string theory remains a hard task. One way to circumvent computing the string partition function in a black hole background is to reduce to a lower dimensional supergravity theory and apply a supersymmetric localization procedure in the near horizon region of the black hole \cite{Dabholkar:2010uh,Dabholkar:2011ec,Dabholkar:2014ema}. This has led to impressive progress and a large body of literature computing exact results for the black hole entropy, see for example \cite{Sen:2023dps,LopesCardoso:2022hvc,Iliesiu:2022kny} for recent discussions on the topic. However, a full understanding of these corrections is still missing due to the fact that the underlying string path integral is unknown and thus measure factors in the localized supergravity path integral remain free parameters. These types of measure factors similarly played a key role in this paper, and are captured in what we call $C(\chi)$, we fixed $C(2)$ using holography. It will therefore be of particular interest to apply our results to worldsheet instanton corrections of the type IIA partition function in four-dimensional black hole backgrounds.\footnote{In \cite{Beasley:2006us} worldsheet instantons were studied in AdS$_2 \times S^2 \times CY_3$ backgrounds of type IIA string theory providing an explicit realization of the OSV conjecture \cite{Ooguri:2004zv,Denef:2007vg}.} Especially for asymptotically AdS black holes where localization procedures have also been attempted \cite{Hristov:2018lod}, but are so far not as well understood. On the other hand, these AdS black holes and their entropy have a direct holographic interpretation in the dual three-dimensional field theory as supersymmetric indices which can be computed using supersymmetric localization. 
Our approach should be directly applicable to determining the non-perturbative contributions to these indices coming from worldsheet instantons and is a subject of future work.

Lastly, we mention that the type IIA limit of the $S^5$ partition function in \eqref{5DLogZ} poses an interesting holographic challenge. The three perturbative terms in the strong coupling expansion should be reproduced by evaluating the supergravity action including higher derivative terms. The leading order term was found by a careful analysis of the holographic renormalization of the two-derivative action in \cite{Bobev:2019bvq}. It would be very interesting to study higher-derivative corrections to eleven-dimensional or seven-dimensional supergravity in order to reproduce the subleading terms in the field theory expression. Together with our results for the series of non-perturbative terms in section \ref{sec:5dsym}, this would constitute a truly remarkable precision test of holography, at leading order in $N$, but exact in $\alpha'$.

\bigskip
\bigskip
\leftline{\bf Acknowledgements}
\smallskip
\noindent We are grateful to Francesco Benini, Nikolay Bobev, Atish Dabholkar, Ziming Ji, Tomoki Nosaka, Pavel Putrov, L{\'a}rus Thorlacius, Arkady Tseytlin and especially Matteo Beccaria for useful discussions. FFG and VGMP are supported by the Icelandic Research Fund under grant 228952-052. FFG and VGMP are partially supported by grants from the University of Iceland Research Fund. JvM is supported by the ERC-COG grant NP-QFT No. 864583 ”Non-perturbative dynamics of quantum fields: from new deconfined phases of matter to quantum black holes”, and by INFN Iniziativa Specifica ST\&FI.

\newpage
\appendix

\section{Measure factors and determinants}
\label{MeasureFactors}
When we compute the one-loop partition function of the string, we start by expanding the action to quadratic order around a given classical configuration. The result is
\be
S_\text{string} \simeq S_\text{cl} + S_\text{FT} + S_\mathbb{K}\,,
\ee
where $S_\text{cl}$ is the classical action of the string and $S_\text{FT}$ is its Fradkin-Tseytlin action. The action of quadratic fluctuations is collected in $S_\mathbb{K}$, which can be denoted by
\be\label{SKdef}
S_\mathbb{K} = \f{1}{4\pi\ell_s^2}\Big(\langle \zeta_a,{\cal K}_{ab}\zeta_b\rangle + \langle \bar\theta_a,{\cal D}_{ab}\theta_b\rangle \Big)\,,
\ee
where 
\be
\langle \zeta,\zeta'\rangle = \int \vol_\gamma \zeta \zeta'\,.
\ee
In the action \eqref{SKdef}, the eight physical bosonic modes of the string are denoted by $\zeta_a$, and the eight fermionic modes are denoted by $\theta_a$. The bosonic operators ${\cal K}_{ab}$ and ${\cal D}_{ab}$ can in principle mix the various string fluctuations. In particular when there is a non-trivial connection, such a mixing occurs. The measure is defined such that
\be
\int [D\zeta]\, \e^{-\f{1}{4\pi\ell_s^2}\langle \zeta,\zeta\rangle} = \int [D\theta]\, \e^{-\f{1}{4\pi\ell_s^2}\langle \bar\theta,\theta\rangle}=1\,.
\ee
This means that the path integral over the one-loop action \eqref{SKdef} reduces to 
\be
\int [D\zeta][D\theta]\, \e^{-S_\mathbb{K}}=(\text{Sdet}\mathbb{K})^{-1/2} \equiv \f{(\det {\cal D}_{ab})^{1/2}}{(\det {\cal K}_{ab})^{1/2}}\,.
\ee
When we have zero modes of either the bosonic operators or fermionic ones, the one-loop action vanishes for those modes, and they have to be treated differently. Usually, we have to carry out an integral over the zero-mode collective coordinates over their target space. For bosonic zero-modes, this gives the volume of the zero-mode target space, whereas for fermionic zero modes, the integral vanishes, which reduces the entire partition function to zero. In this paper, we encounter a combination of bosonic zero modes with a non-compact target space (therefore giving infinite volume factors), together with fermionic zero modes. Regulating these zero modes in a consistent way gives a finite answer for the combined zero-mode partition function, even though individually they would make the partition function diverge or vanish. The conclusion is that whenever we encounter zero modes, we must treat them separately from the non-zero modes, and so we more accurately write
\be
\int [D\zeta][D\theta]\, \e^{-S_\mathbb{K}}=(\text{Sdet}'\mathbb{K})^{-1/2}Z_\text{zero-modes}\,,
\ee
where the partition function of the zero modes has been separated out, leaving only a determinant over the non-zero modes indicated with a prime on the  $\text{Sdet}$.

In this discussion, we have not considered additional measure factors that arise in string theory. These factors are not as well-understood as the field theory contribution discussed here, and are difficult to calculate precisely. However, we can account for these factors by collecting them into a single quantity, denoted as $C(\chi)$. To determine $C(\chi)$, we compare the string theory calculation to the corresponding calculation in the dual field theory, which allows us to fix its value.

\section{Eigenspinors and the spectrum of fermionic operators}\label{App: Fermionic monopole harmonics}
In this appendix we compute the determinant of the kinetic operators encountered for the fermions
\be
{\cal D} = i\slashed{D}+m_1\sigma_3+m_2\,,
\ee
where $\slashed{D} = \slashed{\nabla} - i q \slashed{A}$, $\slashed{\partial} = \sigma_1\partial_\theta + \sigma_2\partial_\varphi$ and $\sigma_{1,2,3}$ are the standard Pauli matrices. This appendix relies heavily on the results of \cite{Wu:1976ge,Wu:1977qk,Dray:1984g} but we refer also to more recent papers\cite{Camporesi:1995fb,Borokhov:2002ib,Benini:2012ui,Pufu:2013vpa} where a similar problem was analyzed. In order to match to the conventions used in e.g. \cite{Dray:1984g}, we change a gauge for the magnetic potential $A$ such that it reads
\be
A = -\f12 \cos\theta\,\dd\varphi\,.
\ee
The fermionic operators act on two-component spin-1/2 fermions and are defined in terms of the curved metric on $S^2$ with length scale $L$
\be
\dd s_2^2 = L^2 (\dd\theta^2 + \sin^2\theta\,\dd\varphi^2)\,.
\ee
In order to eliminate the dependence on the length scale $L$ we compute $\det (L{\cal D})$. Furthermore, in order to obtain a positive spectrum for ${\cal D}$ we always compute the determinant of $\sigma_3$ times ${\cal D}$ in what follows.

In order to determine the spectrum of our operators we first introduce the spin-weighted spherical harmonics or monopole spherical harmonics (see \cite{Dray:1984g}). These are generalizations of the standard spherical harmonics to functions that transform as $f\mapsto \e^{i s\varphi} f$ under rotations around the poles. Here the (half)-integer $s$ is the spin weight of the function $f$. Define the two operators
\be
\begin{split}
\eth^{(s)}  f &= -\sin^{ s}\theta\left[\partial_\theta - i\csc\theta\,\partial_\varphi\right](\sin^{- s}\theta\, f)\ \,,\\
\bar\eth^{(s)}  f &= -\sin^{- s}\theta\left[\partial_\theta + i\csc\theta\,\partial_\varphi\right](\sin^{ s}\theta \, f)\ \,,
\end{split}
\ee
which act on the spin weighted function. We will often omit writing the explicit spin weight where it can be inferred from the context. First, note that the $\eth$ operators satisfy the commutation relation
\be\label{commutator}
\bar\eth^{(s+1)}\eth^{(s)}-\eth^{(s-1)}\bar\eth^{(s)} = sRL^2=2s\,,
\ee
where $R=2/L^2$ is the Ricci scalar on $S^2$. The commutation relation \eqref{commutator} implies that the operators $\eth$ and $\bar\eth$ act as raising and lowering operators on the spin weight $s$. The spin-weighted spherical harmonics ${}_sY_{lm}$ (for integer $s$) are obtained by raising and lowering the spin weight of the standard spherical harmonics ${}_0Y_{lm}\equiv Y_{lm}$ using
\be
\begin{split}
\eth \Yslm{s} &=\sqrt{(l-s)(l+s+1)} \, \Yslm{s+1}\,,\\
 \bar\eth \Yslm{s} &=-\sqrt{(l+s)(l-s+1)} \, \Yslm{s-1}\,.
\end{split}
\ee
The spin weighted spherical harmonics are simultaneous eigenfunctions of $\eth\bar\eth$, $\bar\eth\eth$, ${\bf L}^2$, and $L_z$ which all commute.\footnote{The explicit form of ${\bf L}$ can be found in \cite{Dray:1984g}.} The eigenvalues are given by
\be\label{eigenvalues}
\begin{split}
\eth\bar\eth\Yslm{s} &= (l+s)(l-s+1)\Yslm{s}\,,\\
\bar\eth\eth\Yslm{s} &= (l-s)(l+s+1)\Yslm{s}\,,\\
{\bf L}^2 \Yslm{s} &= l(l+1)\Yslm{s}\,,\\
L_z \Yslm{s} &= m\Yslm{s}\,.
\end{split}
\ee
By construction $l\ge 0$ and $-l\le m \le l$. However the spin weight $s$ puts a further restriction on $l$ since $\eth \Yslm{l} = \bar\eth \Yslm{-l}=0$ which indicates that $l\ge |s|$. The spin weighted spherical harmonics form a complete orthonormal basis for each $s$ in this range
\be
\f{1}{L^2}\int_{S^2}\sqrt{g} \Yslm{s}\,{}_sY_{l'm'} = \delta_{ll'}\delta_{mm'}\,.
\ee

Returning to the original problem, it is now a simple exercise to verify that
\be
L{\cal D} = \begin{bmatrix}(m_1+m_2)L& -i\eth^{(s-1)}\\-i\bar\eth^{(s)}&-(m_1-m_2)L\end{bmatrix}\,,
\ee
where 
\be\label{sdef}
s = \f12 - \f{q}{2}\,.
\ee
The spin weight of the two component functions of the spinor is therefore $s$ and $s-1$. Similarly we find
\be\label{Dsquared}
L^2\tilde{\cal D}{\cal D} = \begin{bmatrix}-\eth \bar\eth + M_f^2L^2& 0\\0&-\bar\eth \eth+ M_f^2L^2\end{bmatrix}\,.
\ee
This shows that the spectral problem for our fermionic operator ${\cal D}$ can be translated to understanding the spectral problem for $\eth \bar\eth$ and $\bar\eth \eth$. In particular, the eigenspinors of the squared fermionic operator \eqref{Dsquared} can be expressed in terms of the spin-weighted spherical harmonics 
\be
\psi_{lm} = \begin{bmatrix}\Yslm{s}\\\Yslm{s-1}\end{bmatrix}\,.
\ee 
These spinors are eigenspinors also of the linear operator ${\cal D}$ but not in the usual sense as they do not possess standard eigenvalues, but rather eigenmatrices:
\be\label{Dpsi}
L{\cal D}\psi_{lm} = \begin{bmatrix}(m_1+m_2)L& -i\sqrt{(l-s+1)(l+s)}\\i\sqrt{(l-s+1)(l+s)}&-(m_1-m_2)L\end{bmatrix}\psi_{lm}\,.
\ee
Because of the form of the operator \eqref{Dsquared} and the fact that $\eth \Yslm{l} = 0$, and $\bar\eth \Yslm{-l}=0$ we see that the smallest possible value of $l$ is either $l=-s$ or $l=s-1$ depending on which is positive. Using \eqref{sdef} we see that this means
\be\label{inequality}
l+\f12 \ge\f{|q|}{2}\,.
\ee
Furthermore, the expression \eqref{Dpsi} is invalid for these cases since one component of the eigenspinor is equal to zero. Before analysing these threshold cases we note that for $l+1/2 > |q|/2$ the eigenvalue of $L^2\tilde{\cal D}{\cal D}$ is given by
\be
(l+1/2)^2 -\f{q^2}{4}+M_f^2L^2\,,
\ee
which can be observed to also equal the determinant of the eigenmatrix for $L{\cal D}$ associated with the eigenspinor $\psi_{lm}$ (recall that we include the third Pauli matrix when computing the determinant of ${\cal D}$).

When the inequality \eqref{inequality} is saturated, we have to analyse the problem separately. Let $q>0$ and $l+1/2 = q/2$, then $s=-l$. This of course means that ${}_{s-1}Y_{lm}$ does not exist as noted before. The eigenspinor is therefore simply given by 
\be
\begin{bmatrix}{}_{-l}Y_{lm}\\0\end{bmatrix}\,,
\ee
with
\be
L{\cal D} \begin{bmatrix}{}_{-l}Y_{lm}\\0\end{bmatrix} = (m_1+m_2)L\begin{bmatrix}{}_{-l}Y_{lm}\\0\end{bmatrix}\,,
\ee
and the eigenvalue is clearly read to be $(m_1+m_2)L$. Note that the degeneracy of this mode is still given by $2l+1=q=|q|$ as usual. For the opposite sign of $q$, namely $q<0$ and $l+1/2 = -q/2$, then $s=l+1$ and the eigenfunction is now
\be
\begin{bmatrix}0\\{}_{l}Y_{lm}\end{bmatrix}\,,
\ee
with eigenvalue of $\sigma_3L{\cal D}$ given by $(m_1-m_2)L$. Once again the degeneracy is given by $2l+1=-q=|q|$.

Let us now collect this information and compute the determinant of ${\cal D}$. Collecting the two cases we can write
\be\label{logdetqpos}
\log\det (L{\cal D}) = \sum_{l+1/2 = |q|/2+1}^\infty(2l+1)\log\Big[(l+1/2)^2 -\f{q^2}{4}+M_f^2L^2\Big] + |q|\log L(m_1+\text{sgn}(q)m_2)\,,
\ee
where $\text{sgn}(q)$ denotes the sign of $q$. Notice that the mode that we are counting separately would be a zero-mode if we would have massless fields. To make contact with the discussion in the main text we should adjust our expression \eqref{logdetqpos} such that the sums start from $|q|/2$ and not $|q|/2+1$. We can remedy this by adding and subtracting terms 
\be
\log\det (L{\cal D}) = \sum_{l = |q|/2}^\infty(2l)\log\Big[l^2 -\f{q^2}{4}+M_f^2L^2\Big] - |q|\log L(m_1-\text{sgn}(q)m_2)\,,
\ee
where we have also reindexed the sum $l+1/2\to l$.

\section{Holographic duals to mass deformed ABJM}\label{App: Holographic duals to mass deformed ABJM}
\label{11DsolsandSU3}
\subsection{\texorpdfstring{$\SO(4)\times\SO(4)$}{SO4xSO4} deformed background in 11 dimensions}
The eleven-dimensional solution dual to the $\SO(4)\times\SO(4)$ mass deformation of ABJM is obtained by uplifting the four-dimensional supergravity solution of \cite{Freedman:2013ryh} by using the formulae in \cite{Cvetic:1999au}. The resulting metric and three-form take the form
\be
\begin{split}
\dd s_{11}^2 &=   \f{N^{2/3}\pi^{2/3}\ell_{pl}^2}{(2\lambda)^{1/3}}(Y_1 Y_2)^{1/3}\bigg(\dd s_{\text{AdS}_4}^2 + 4\dd\theta^2 + \f{\cos^2\theta}{Y_1}\dd s_{S^3_1}^2+\f{\sin^2\theta}{Y_2}\dd s_{S^3_2}^2\bigg)\,,\\
A_3 &=-\f{N i\pi \ell_{pl}^3}{\sqrt{2\lambda}} \bigg(3U \vol_{S^3}\mp\f{c }{\cosh^2(\rho/2)} \Big( \f{\cos^4\theta}{Y_1}\vol_{S^3_1} +\f{\sin^4\theta}{Y_2}\vol_{S^3_2} \Big)\bigg)\,,
\end{split}
\ee
where the metric on AdS$_4$ is Euclidean, round and in spherical coordinates, reflecting the fact that the field theory is living on $S^3$. The transverse directions also contain two three-spheres which we denote by $S^3_1$ and $S^3_2$.  The deformation of the standard AdS$_4\times S^7$ geometry is implemented by the three functions
\be
\begin{split}
Y_1&=1+c \f{\cos^2\theta}{\cosh^2(\rho/2)}\,,\\
Y_2&=1-c \f{\sin^2\theta}{\cosh^2(\rho/2)}\,,\\
U &= \f{1}{12}\Big( -9\cosh\rho + \cosh(3\rho) + 16 c \cos(2\theta)\sinh^4(\rho/2) \Big)\,.
\end{split}
\ee
The signs in the three-forms $A_3$ reflects the two branches of solutions discussed in more detail in the main text. We can map between the two branches by performing the simultaneous transformation of the coordinate $\theta$ and reversing the sign of $c$
\be
c\mapsto-c\,,\qquad \theta\mapsto \theta+\f\pi2\,.
\ee

In order to reduce to ten-dimensions we must reduce over the hopf fiber of $S^7$. In our coordinate system the Hopf fiber of $S^7$ is a diagonal combination of the Hopf fibers of $S_1^3$ and $S_2^3$, this coordinate is furthermore restricted in its range to be $2\pi/k$-periodic. Explicitly reducing over this circle, identifying the eleven-dimensional Planck length $\ell_{pl}$ with the string length $\ell_s$, and adding the correct period to the $B_2$ field to account correctly for the ABJ theory as described in the main text, results in the ten-dimensional solution in section \ref{ssec:singlemassdeform}.

\subsection{\texorpdfstring{$\SU(3)$}{SU3} deformed background in type IIA}
A second simple mass deformation of ABJM theory is the equal mass deformation which exhibits $\SU(3)$ enhanced symmetry. In order to study non-perturbative corrections to the free energy of the QFT we should study the worldsheet instantons in the dual ten-dimensional deformed geometry. In this appendix we take the first step on this path by constructing the ten-dimensional geometry dual to the particular mass deformation in question. This is found simply by uplifting the analytic four-dimensional supergravity solution in \cite{Freedman:2013ryh} using the uplift formulae in \cite{Pilch:2015dwa}. We will not go further in this paper in studying worldsheet instantons but leave that as future work.

For notational simplicity we introduce the following two functions
\begin{equation}
	S = \frac{c(c-1)(1-\rho^2)(1+c^3 \rho^2)}{(1+c^3)(1+c^3 \rho^4)}\,,\qquad C=1-\frac{2c^3(1-\rho^2)^2}{(1+c^3)(1+c^3\rho^4)}\,,
\end{equation}
where $\rho$ is the radial holographic coordinate, and $c$ is related to the field theory mass deformation through
\begin{equation}
	\delta = \frac12 \frac{c(1+c)}{1+c^3}\,.
\end{equation}
The ten-dimensional metric then becomes
\begin{equation}
	\rmd s^2 = L^2 \sqrt{1+ S^2 \sin^2 \theta } \left[ \rmd s_{4}^2 + \frac{1}{C+S} \rmd s_6^2 \right]\,,
\end{equation}
where 
\begin{equation}\label{SU3metric}
	\begin{aligned}
		\rmd s_4^2 =&\, \frac{4(1+c^3)(1+c^3 \rho^4)}{(1-\rho^2)^2(1+c^3 \rho^2)^2}(\dd\rho^2 + \rho^2 \dd\Omega^2)\,,\\
		\rmd s_6^2 =&\, \rmd \theta^2  +  \frac{16\cot^2 \frac{\theta}{2}}{1+ S^2 \sin^2 \theta} \omega^2 + \frac{4(C+S)\sin^2\frac{\theta}{2}}{C + S \cos\theta} \rmd s_{\CP^2}^2\,.
	\end{aligned}
\end{equation}
Here we have introduced 
\be
\omega = \f12 \sin^2\f\theta2(\dd\varphi + \omega_{\CP^2})\,,
\ee
which is related to the K\"ahler form on $\CP^3$ through $J = \dd \omega$. In a similar fashion, the K\"ahler form on $\CP^2$ is related to $\omega_{\CP^2}$ via $J_{\CP^2} = \dd\omega_{\CP^2}$. The metric $\rmd s_{\CP^2}^2$ on $\CP^2$ appearing in equation \eqref{SU3metric} is the Fubini-Study metric of radius one.
The dilaton and $B_2$ field are
\begin{equation}
	\e^{2\Phi} = \f{\pi(2\lambda)^{5/2}}{N^2}\frac{(1+S^2\sin^2\theta)^{3/2}}{(C+S)(C+S\cos\theta)^2}\,,\qquad B_2 =  \frac{4\pi\rmi \ell_s^2\sqrt{2\lambda} S}{C+S\cos\theta} \frac{c+1}{c-1}  \left[ J - \frac{S \sin \theta}{C+S} \rmd \theta \wedge \omega  \right]\,.
\end{equation}
Finally, the RR gauge potentials are given by
\begin{equation}
	\begin{aligned}
		C_1 =&\, \frac{2N\ell_s}{\lambda} \frac{1 + 2 S(C+S) \cos^2 \frac{\theta}{2}}{1 + S^2 \sin^2 \theta } \omega\,,\\
		C_3^{\text{ext}} =&\, - \frac{2N\ell_s}{\lambda} B_2 \wedge \omega\,,\\
		C_5^{\text{ext}} =&\, - 32 \pi^2  N\ell_s^5 \frac{1 + 2 S(S-C) \sin^2 \frac{\theta}{2}}{(C+S\cos\theta)^2} J \wedge J \wedge \omega\,.
	\end{aligned}
\end{equation}
Notice that we have given the gauge potential $C_5$ and $C_3$ in the $\CP^3$ directions instead of giving the full $C_3$ potential which would also have terms along the deformed AdS$_4$ directions. The type IIA field strengths can be computed from these using 
\begin{equation}
	\begin{aligned}
		H_3 &= \dd B_2\,,\\
		F_2 &= \dd C_1\,,\\
	\end{aligned}
	\qquad 
	\begin{aligned}
		F_4^\text{ext} &= \dd C_3^\text{ext} - H_3\w C_1\,,\\
		F_6^\text{ext} &= \dd C_5^\text{ext} - H_3\w C_3^\text{ext}\,,
	\end{aligned}
\end{equation}
and
\begin{equation}
	F_4 = F_4^{\text{ext}} - \rmi \star F_6^{\text{ext}}\,.
\end{equation}

With this background it is a simple matter to uplift the solution to eleven dimensions using the standard uplift formulae (see e.g. \cite{Blumenhagen:2013fgp}).

\newpage
\bibliography{refs}
\bibliographystyle{JHEP}

\end{document}